\documentclass[journal]{IEEEtran}

\IEEEoverridecommandlockouts

\usepackage[utf8]{inputenc} 

\usepackage[T1]{fontenc}
\usepackage{url}
\usepackage{ifthen}
\usepackage{cite}
\usepackage{overpic}
\usepackage[cmex10]{amsmath} 

\usepackage{textcase}
\usepackage[tablename=Table]{caption}
\usepackage{blindtext}
\usepackage{floatrow}
\usepackage{soul}
\usepackage{changes}
\usepackage{gensymb}
\usepackage{cite}
\usepackage{amsmath,amssymb,amsfonts}

\usepackage{textcomp}
\usepackage{graphicx}

\usepackage{amssymb}
\usepackage{amsfonts}
\usepackage{amsmath}
\usepackage{epsfig}
\usepackage{epigraph}
\usepackage{color}
\usepackage{fancybox}
\usepackage{textcomp}
\usepackage{multirow}
\usepackage{setspace}
\usepackage{psfrag}
\usepackage{booktabs}
\usepackage{float}
\usepackage[caption = false]{subfig}
\usepackage{algorithm}
\usepackage{algpseudocode}
\usepackage{mathtools, nccmath, bigints, amsfonts}
\usepackage{array}
\usepackage{stfloats}
\newcolumntype{L}{>{\centering\arraybackslash}m{3cm}}

\usepackage{amsthm}
\theoremstyle{definition}
\usepackage{xcolor}
\usepackage{mathrsfs}
\usepackage{tikz}
\usepackage{balance}

\newcommand{\Ttran}{^{\mathsf{T}}}

\newfloatcommand{capbtabbox}{table}[][0.4\textwidth]
 \usepackage[font={small},figurename=Fig.]{caption}

\makeatletter 
\pretocmd\@bibitem{\color{black}\csname keycolor#1\endcsname}{}{\fail}
\newcommand\citecolor[1]{\@namedef{keycolor#1}{\color{black}}}
\makeatother

    \def\Complex{{\rm\rule[.23ex]{.03em}{1.1ex}\kern-.3em{C}}}

    \newcommand{\be}{\begin{equation}} \newcommand{\ee}{\end{equation}}
    \newcommand{\bea}{\begin{eqnarray}} \newcommand{\eea}{\end{eqnarray}}
    \newcommand{\benum}{\begin{enumerate}} \newcommand{\eenum}{\end{enumerate}}

\def\Htran{\mbox{\tiny $\mathrm{H}$}}
\def\Ttran{\mbox{\tiny $\mathrm{T}$}}

\def\imaginary{\mathsf{j}} 
\def\Htran{\mbox{\tiny $\mathrm{H}$}}
\def\Ttran{\mbox{\tiny $\mathrm{T}$}}

\begin{document}

\title{Joint Channel Scanning and Geometry-Aware Codebook Design for Movable Antenna Arrays}
\author{Amna~Irshad,~Luca~Antonelli,~Vitaly~Petrov,~Luca~Sanguinetti, and Emil~Bj{\"o}rnson\\
\thanks{A.~Irshad, V.~Petrov, and E.~Bj{\"o}rnson are with the Department of Communication Systems,  KTH Royal Institute of Technology, Stockholm, Sweden. ~Luca~Antonelli and~Luca~Sanguinetti are with the Dipartimento di Ingegneria dell’Informazione, University of Pisa. Email:~\{amnai,emilbjo,vitalyp\}@kth.se; luca.antonelli@phd.unipi.it, luca.sanguinetti@unipi.it}%
\thanks{This work was supported by the Grant 2022-04222 from the Swedish Research Council.}

}

\maketitle

\begin{abstract}
Multiple-input multiple-output (MIMO) systems rely on spatial degrees of freedom, but a fixed antenna geometry cannot adapt to the propagation environment. Movable antenna (MA) systems address this by letting antenna positions adapt to the channel. In a discrete MA architecture, the active antennas are selected from a large set of candidate positions, which requires channel state information (CSI) at the candidates. With a limited number of radio-frequency (RF) chains, this CSI must be acquired through successive pilot rounds, a process we call antenna scanning, whose overhead can offset the benefits of MA reconfiguration. This paper studies discrete MA geometry optimization when only a fraction of the candidate positions is scanned before the antenna configuration is chosen. We first show that a geometry-aware (GA) steering-vector codebook, built from the actual antenna coordinates, reconstructs the channel more accurately than a conventional discrete Fourier transform (DFT) codebook once the optimized array departs from a uniform grid. We then characterize how the achievable sum rate varies with the scanned fraction, showing that the fraction needed to approach full-scan performance depends strongly on the CSI acquisition method: least-squares (LS) estimation degrades sharply at low scanning fractions, whereas codebook-based acquisition remains robust under much sparser scanning. Building on this, we propose an adaptive scanning procedure that probes candidates in batches and stops automatically based on the acquired pilots, so the scanning budget need not be fixed in advance. Finally, we study user mobility, quantifying how much sum rate is lost as users move away from their original positions and how much of this loss is recovered by re-scanning and re-selecting the antenna geometry, compared with keeping the aperture fixed with outdated or fresh CSI. Numerical results show that adaptive scanning retains approximately $99.5\%$ of the full-scan sum rate while scanning as little as $11\%$ of the candidate positions.

\end{abstract}


\section{Introduction}

The rapid evolution of wireless applications, including extended reality, ultra-high-definition multimedia, and massive machine-type communications, is driving the demand for higher spectral efficiency, reliable communication, and enhanced connectivity in future wireless systems, including 6G. Multiple-input multiple-output (MIMO) technology exploits the spatial degrees of freedom provided by multiple antennas to improve spectral efficiency and communication reliability \cite{AKBAR2025104040}. In principle, the MIMO technology can evolve by increasing the number of antennas and deploying more base stations (BSs), but this might not be financially sustainable. This raises the question of whether one can achieve better performance with a fixed number of antennas and a fixed BS grid \cite{Bjornson2026a}. Conventional MIMO arrays have a given number of antennas at fixed antenna positions, limiting their ability to adapt their spatial configuration to changing propagation conditions and user locations~\cite{Bjornson2019c}. Although digital beamforming can adjust the signals' amplitudes and phases to shape the effective beam pattern, it cannot change the physical antenna positions or directly reconfigure the channel by moving the antennas. Consequently, the spatial configuration of a fixed array may not be optimal for a given propagation environment or user distribution~\cite{MAtutorial}.

Movable antenna (MA) systems have recently emerged as a promising solution to the above challenges. By design, MA-enabled systems feature an additional spatial degree of freedom, as the antenna positions can be adapted to the propagation environment \cite{fluid21,movgeneral}. This allows the channel itself to be reshaped, rather than merely accommodated: (i) channel gain can be improved by moving antennas away from deep fades, which enhances received signal strength and diversity; (ii) channel rank and conditioning can be improved by increasing the number of effective spatial modes and reducing the singular value spread (i.e., the condition number) of the channel matrix, which increases spatial multiplexing gain; and (iii) array response can be reshaped to improve beamforming, e.g., by steering nulls toward interferers and suppressing sidelobes\cite{xiao2023multiuser,beam23,antenna24,PhaseMA}. Closely related fluid antenna systems (FAS) exploit reconfigurability through position or port selection \cite{wong2020performance,REMAA25,FAfinite}. MA formulations fall into two main categories. In the continuous model, each antenna can occupy any position within a given region \cite{MADMA,MABF}. In the discrete model, the region is divided into a finite set of candidate locations, to which the available RF chains are moved or activated \cite{discreteMA,discreteMA2}. In both cases, the search space grows rapidly with the aperture size, and for large apertures the number of candidate antenna positions can reach the hundreds.

Selecting active antenna positions from a discrete candidate set dynamically is a form of antenna selection \cite{Antennasel2004,Sanayei2004}. Although an exhaustive search over all possible antenna location subsets is generally computationally prohibitive (e.g., placing 16 antennas from 64 candidate locations already yields hundreds of \emph{trillions} of possible configurations), even low-complexity selection algorithms require channel state information (CSI) to identify promising antenna configurations. Acquiring this CSI is challenging because the number of RF chains equals the number of active antennas, which is much smaller than the number of candidate positions. We refer to measuring the channel between the users and an antenna placed (or switched) at a given candidate position as \emph{scanning} that position, and to repeating this over the candidate set as \emph{antenna scanning}. Unlike conventional channel estimation, which estimates the channel of the fixed, active antennas, antenna scanning determines which antenna positions are worth activating. Its overhead grows with the number of candidates, which can diminish the practical benefits of MA reconfigurability.

Importantly, however, the availability of a large candidate set does not necessarily imply that every candidate location must be scanned at each point in time to obtain good antenna-selection performance. If a carefully chosen subset of locations provides sufficient information to identify a high-performing antenna configuration, partial scanning could substantially reduce acquisition overhead while retaining most of the achievable sum rate. Existing MA and antenna-selection works typically assume that full CSI of the candidate set is available, which leaves open a fundamental question: \emph{How much of the candidate region must be scanned to approach full-scan performance, and how does the required scanning effort depend on the CSI acquisition method and the spatial distribution of the scanned locations?}

The effectiveness of partial scanning depends on how the measurements collected at the probed locations are used to characterize the channel across the candidate region. Conventional estimators obtain channel information directly at the scanned positions and require no assumption on the channel structure beyond its statistics (e.g., least-squares (LS)). In contrast, codebook-based acquisition represents the channel using a set of steering vectors~\cite{DFT2006,codebook2014}, whereas compressed-sensing approaches recover the unobserved channel components by exploiting the spatial sparsity of the channel, i.e., the small number of dominant propagation paths~\cite{Comp23,xiao2024framework,compMA}. However, the standard DFT codebook, designed for a uniformly spaced array~\cite[Ch. 7]{Tse_Viswanath_2005}, may not accurately represent the spatial response of an optimized, irregular MA geometry. A geometry-aware (GA) codebook constructed from the antenna coordinates can address this mismatch. These differences motivate evaluating not only the number of scanned locations but also how the scanning fraction interacts with the channel estimation approach.

This problem has been partially studied. The design of the codebook for MA arrays was considered in \cite{codebookMA} to improve the beamforming gain of hierarchical codewords in the presence of noise, and in \cite{nearMA}, which develops a codebook to reduce training overhead by an MA-assisted near-field system serving a single user. Both works treat the antenna geometry as a codebook design variable or focus on beam training for a one-dimensional MA array; they do not quantify how much of a candidate region must be scanned for antenna selection, do not reconstruct the channel on an irregular selected geometry, and do not consider mobile users for a multi-user MIMO scenario. To the best of our knowledge, no existing study combines (i) codebook-based channel reconstruction on irregular two-dimensional MA geometries, (ii) the scanning effort needed to approach full-scan performance, and (iii) re-scanning under user mobility. To address this, we aim to answer three questions: 
(i) How accurately can a geometry-aware codebook reconstruct the channel of an irregular MA geometry compared with a conventional DFT codebook? (ii) How many candidate locations must be scanned before the achievable sum rate saturates, how does this saturation point depend on the channel estimation approach (LS or codebook-based), and does the spatial distribution of the scanned locations matter? (iii) When users move, how much scanning effort is needed to recover the rate, and can it be substantially reduced by seeding the scan with the previously optimized geometry rather than starting from a full scan?

\subsection{Main Contributions}

The main contributions of this paper are:

\begin{itemize}

\item \textbf{Geometry-aware codebook construction and evaluation:}
We formulate a GA codebook by evaluating steering vectors at the selected movable-antenna coordinates and benchmark it against physical-grid and virtual-array DFT codebooks. This comparison shows how matching the codebook to the antenna geometry affects channel reconstruction and downlink sum-rate performance.

\item \textbf{Characterization of scanning-performance tradeoffs:}
We quantify the relationship between scanning effort and achievable sum rate for LS estimation and codebook-based channel reconstruction. By examining different scanning fractions and spatial sampling strategies, we identify the acquisition regimes in which additional scanning yields diminishing rate improvements and characterize how these regimes depend on the channel estimation approach.

\item \textbf{Practical energy-based adaptive scanning and mobility-aware performance evaluation:}
To avoid relying on the unknown true channel over the full candidate region, we develop and evaluate a sequential scanning strategy that uses an energy-based metric computed from acquired channel estimates to guide its stopping decision. The strategy expands the scanned set in batches and stops when the relative change in the metric becomes sufficiently small. We quantify its scanning overhead and evaluate the resulting sum-rate performance against fixed-budget scanning strategies. We further evaluate performance under user displacement by comparing fixed-geometry operation, renewed antenna selection, and antenna subsets optimized at the original user positions, under both fresh and stale CSI, and quantifying the tradeoffs among scanning overhead, channel variation, and downlink sum rate.

\end{itemize}

\textit{Outline:} Section~\ref{sec:system_model} presents the system model, including the discrete MA architecture, pilot-based channel acquisition, and the downlink transmission model and performance metric. It also introduces the spatial codebook representations (the virtual-array DFT, physical-grid DFT, and geometry-aware steering-vector codebooks) and examines their sensitivity to irregular antenna positions. Section~\ref{sec:optimization} formulates the discrete antenna geometry optimization problem and details the antenna selection steps. Section~\ref{sec:fullscanning} evaluates codebook performance under full spatial scanning. Section~\ref{sec:scanning} develops the scanning-aware optimization, analyzes partial scanning, introduces adaptive scanning, and examines user mobility. Section~\ref{Sect_conclusion} concludes the paper.

\textit{Notation:} Scalars, vectors, and matrices are denoted by lowercase, bold lowercase, and bold uppercase letters, respectively. The Hermitian transpose is denoted by $(\cdot)^{\Htran}$ and expectation by $\mathbb{E}\{\cdot\}$. The transpose and imaginary unit are denoted by $(\cdot)^{\Ttran}$ and $\imaginary$, respectively. For a complex scalar $x$, $|x|$ denotes its modulus and $|\cdot|^2$ the squared Euclidean norm. The notation $a \triangleq b$ denotes equality by definition. For a matrix $\mathbf X$ and an index set $\mathcal I$, $[\mathbf X]_{\mathcal I,:}$ denotes the submatrix formed by the rows of $\mathbf X$ indexed by $\mathcal I$ (all columns retained). 

\section{System Model}\label{sec:system_model}

\begin{figure*}[t!]
    \centering
   \includegraphics[width=1\textwidth]{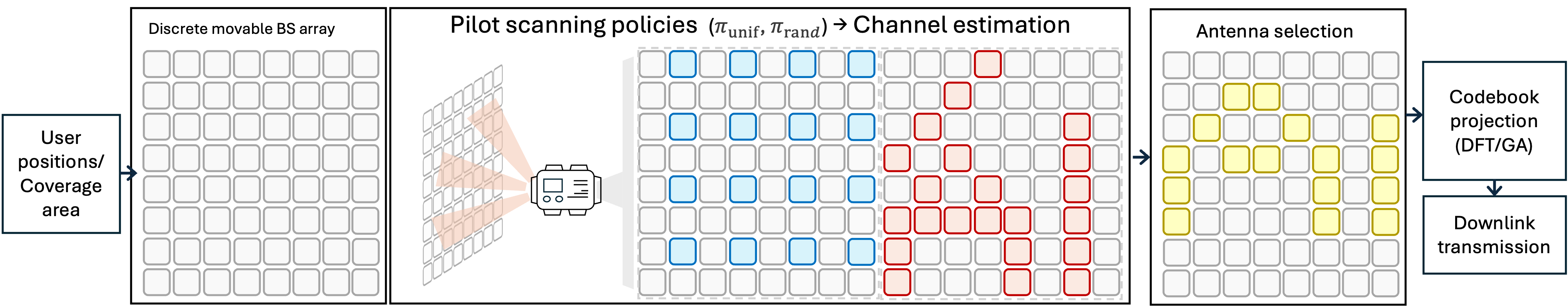}
    \caption{System model and operational flow of the proposed discrete movable antenna array for multi-user downlink transmission.}
    \label{fig:system_flow}
\end{figure*}

We consider a multi-user downlink system in which a BS equipped with a movable antenna array communicates with $K$ single-antenna users. The system operates at carrier frequency $f_c$ with wavelength $\lambda=c/f_c$ over a narrowband channel. Unlike continuously movable antenna architectures \cite{MAtutorial}, each antenna element is constrained to occupy one of a finite number of candidate locations. This discrete array architecture captures practical implementations in which antennas are repositioned among predefined mechanical or switching locations (e.g., via stepper motors). It shares with fluid-antenna port selection \cite{FAS} and pinching-antenna activation \cite{PA} the idea of choosing active radiating positions from a finite candidate set, but here each of the $M$ RF chains is tied to one physically movable antenna.
The overall system model and signal processing flow, including pilot scanning, channel estimation, antenna selection, codebook projection, and downlink transmission, are shown in Fig.~\ref{fig:system_flow}.

\subsection{Discrete Movable Antenna Architecture}
Let the BS have a set of physically feasible candidate locations 
\begin{align}
    \mathcal A_{\mathrm{grid}}=\{1,\ldots,M_{\mathrm{grid}}\}
\end{align}
arranged on a planar $N_y\times N_z$ grid in the $y$-$z$ plane such that $M_{\mathrm{grid}}=N_yN_z$, with $\lambda/2$ spacing so that mutual coupling between antenna elements is negligible. All antenna elements are isotropic and the position of the $i$th candidate is $\tilde{\mathbf a}_i\in\mathbb R^3$. The BS has $M\triangleq M_{\mathrm{RF}}\le M_{\mathrm{grid}}$ RF chains, each driving exactly one antenna element \cite{FASmod}, but the assignment can be changed over time. The RF chains are connected to the candidate positions through a reconfigurable connection, e.g., an $M_{\mathrm{grid}}\times M$ RF switching network or mechanical repositioning of the elements. The connection is fully flexible: any RF chain can be connected to any candidate position. The BS can therefore activate any subset $\mathcal I=\{i_1,\ldots,i_M\}\subset\mathcal A_{\mathrm{grid}}$ with $|\mathcal I|=M$, i.e., one of $\binom{M_{\mathrm{grid}}}{M}$ configurations, and the switching time is neglected. The coordinates of the selected elements are $\mathbf a_m=\tilde{\mathbf a}_{i_m}$, the resulting array geometry is $\mathbf A=[\mathbf a_1,\ldots,\mathbf a_M]\in\mathbb R^{3\times M}$, and the user locations are $\mathbf B=[\mathbf b_1,\ldots,\mathbf b_K]\in\mathbb R^{3\times K}$.

We consider a free-space line-of-sight (LoS) propagation scenario since MAs show the most prominent performance gains in such deployments \cite{Irshad2026a}. The channel coefficient between the $i$th candidate antenna location and the $k$th user is 
\begin{equation}
h_{k,i}=\sqrt{\beta_{k,i}}\exp\!\left(-j\tfrac{2\pi}{\lambda}d_{k,i}\right),
\quad d_{k,i}=\|\mathbf b_k-\tilde{\mathbf a}_i\|,
\end{equation}
where $\beta_{k,i}=\Big(\tfrac{\lambda}{4\pi d_{k,i}}\Big)^2$ represents the free-space path loss. Let $\mathbf H_{\mathrm{grid}}\in\mathbb C^{M_{\mathrm{grid}}\times K}$ with $[\mathbf H_{\mathrm{grid}}]_{i,k}=h_{k,i}$ be the uplink channel matrix over the full grid of antenna candidate locations. We assume TDD operation with perfect reciprocity calibration of all RF chains, so the downlink channel for configuration $\mathcal I$ is
\begin{equation}
\mathbf H(\mathcal I)=\big[\mathbf H_{\mathrm{grid}}\big]_{\mathcal I,:}^{T}
\in\mathbb C^{K\times M}.
\end{equation}
Hence, changing $\mathcal I$ modifies both the array geometry and the channel matrix by selecting which rows of $H_{\mathrm{grid}}$ are utilized. This flexibility is central to the scanning problem considered in this paper. The channel is assumed constant over all pilot batches and the subsequent downlink transmission.

The LoS expression is the ground-truth channel used to generate realizations. However, the BS does not know $\mathbf B$ or $\mathbf H_{\mathrm{grid}}$ a priori, but  must acquire channel information only through uplink pilots. Let $\mathcal S_{\mathrm{scan}}\subseteq\mathcal A_{\mathrm{grid}}$ denote the \emph{scan set}, i.e., the set of candidate locations whose channels are probed by pilots. Only locations in $\mathcal S_{\mathrm{scan}}$ are observed, so the BS can select a configuration $\mathcal I$ only from $\mathcal S_{\mathrm{scan}}$ (i.e., $\mathcal I\subseteq\mathcal S_{\mathrm{scan}}$). For notational simplicity, we first consider \emph{full scanning}, where $\mathcal S_{\mathrm{scan}}=\mathcal A_{\mathrm{grid}}$ and every candidate location is probed to study the effect of channel estimation approaches. The proposed practical implementation with reduced scan sets is then introduced in
Section~\ref{sec:scanning}.

\subsection{Pilot-Based Channel Acquisition}\label{sec:pilot}

The BS probes the locations in $\mathcal S_{\mathrm{scan}}$ (here, $\mathcal S_{\mathrm{scan}}=\mathcal A_{\mathrm{grid}}$, i.e., full scanning). Since the BS has only $M=M_{\rm RF}$ RF chains, it probes at most $M$ candidate locations per pilot batch, so the scan set is covered in
$n_b=\lceil |\mathcal S_{\mathrm{scan}}|/M\rceil$
batches $\mathcal T_1,\ldots,\mathcal T_{n_b}$ of at most $M$ contiguous locations (row-major order). For batch $\mathcal T_b$, let $\mathbf H_b=[\mathbf H_{\mathrm{grid}}]_{\mathcal T_b,:}\in\mathbb C^{|\mathcal T_b|\times K}$. The received pilot signal is
\begin{equation}
\mathbf Y_b=\sqrt{P_{\mathrm{ul}}}\,\mathbf H_b\boldsymbol\Phi+\mathbf N_b,
\end{equation}
where the pilot matrix $\boldsymbol\Phi\in\mathbb C^{K\times\tau_p}$ satisfies $\boldsymbol\Phi\boldsymbol\Phi^H=\tau_p\mathbf I_K$ (requiring $\tau_p\ge K$) and $\mathbf N_b$ has i.i.d.\ $\mathcal{CN}(0,\sigma_n^2)$ entries, with noise variance $\sigma_n^2$. The LS estimate is
\begin{equation}
\widehat{\mathbf H}_b=\frac{\mathbf Y_b\boldsymbol\Phi^H}{\tau_p\sqrt{P_{\mathrm{ul}}}}
=\mathbf H_b+\tilde{\mathbf N}_b,
\end{equation}
where the effective noise matrix  $\tilde{\mathbf N}_b$ has i.i.d. $\mathcal{CN}\big(0,\sigma_n^2/(\tau_pP_{\mathrm{ul}})\big)$ entries. Stacking all batches gives the full-grid estimate $\widehat{\mathbf H}_{\mathrm{grid}}=\mathbf H_{\mathrm{grid}}+\tilde{\mathbf N}$. For a configuration $\mathcal I$, the BS extracts $\mathbf Z(\mathcal I)=[\widehat{\mathbf H}_{\mathrm{grid}}]_{\mathcal I,:} =[\mathbf z_1,\ldots,\mathbf z_K]\in\mathbb C^{M\times K}$.                                                            

\subsection{Spatial Codebook Representation}\label{sec:codebook}

The LS estimate for the $k$th user will most likely not be a plausible LoS channel. To reduce estimation error, the BS can project the estimate onto a spatial codebook to denoise the estimate. Let $\mathbf W(\mathcal I)=[\mathbf w_1,\ldots,\mathbf w_{Q^2}]\in\mathbb C^{M\times Q^2}$ denote a codebook of unit-norm columns, where $M$ is the number of active antennas and $Q$ is the number of candidate spatial-frequency samples per axis. The codebook samples the two spatial frequencies $(u_y,u_z)\in[-1,1]^2$, i.e., the direction cosines along the $y$- and $z$-axes of the array plane, so it contains $Q^2$ codewords. Each column $\mathbf w_q$ is a candidate steering vector, i.e., the array response of the selected elements $\mathcal I$ to one candidate direction.  Increasing $Q$ provides denser spatial-frequency sampling at the cost of a larger codebook and increased search complexity.
For the $k$th user, the BS first finds the codebook vector best aligned with the LS estimate, \eqref{eq:proj_index}, and then projects the estimate onto it, \eqref{eq:proj_est}
\begin{align}
q_k^\star &= \arg\max_{q}\left|\mathbf w_q^H\mathbf z_k\right|^2, \label{eq:proj_index}\\
\hat{\mathbf h}_k &= \left(\mathbf w_{q_k^\star}^H\mathbf z_k\right)\mathbf w_{q_k^\star}. \label{eq:proj_est}
\end{align}
Stacking the projected vectors gives $\widehat{\mathbf H}_{\mathrm{cb}}(\mathcal I)=[\hat{\mathbf h}_1,\ldots,\hat{\mathbf h}_K]^T\in\mathbb C^{K\times M}$, the denoised channel estimate used for precoder design.

In this paper, we consider three codebooks that differ in how they account for antenna geometry. A user in the far-field produces, across the $M$ active antennas, a phase progression determined only by its direction, so projecting the LS estimate onto the best-matching steering vector keeps the LoS component and removes the noise orthogonal to it. The difficulty is that the active set $\mathcal I$ is arbitrary, so the active array is generally irregular and changes with $\mathcal I$. The codebooks differ in how they handle this. VA-DFT ignores the antenna positions and serves as a baseline; PG-DFT uses the physical grid positions but a fixed set of directions; GA uses the actual antenna coordinates and a freely chosen set of directions.

\subsubsection{Virtual-Array DFT (VA-DFT) Codebook}
The baseline is the conventional DFT codebook, which assumes the $M$ active antennas form a compact uniform rectangular array. We call it \emph{virtual} because this array does not exist physically in our setup: it is an imagined $\sqrt M\times\sqrt M$ array with the same number of elements as the active set.
The VA-DFT codebook is
\begin{equation}
\mathbf W_{\mathrm{VA}}=\mathbf F_v\otimes\mathbf F_v
\in\mathbb C^{M\times Q^2},
\end{equation}
where
\begin{equation}
[\mathbf F_v]_{n,q}
=\frac{1}{M^{1/4}}
e^{-j2\pi(n-1)(q-1)/Q},
\qquad n=1,\ldots,\sqrt M.
\end{equation} 
The active antennas are sorted by their $z$-coordinates, the sorted list is partitioned into $\sqrt M$ groups, and each group is sorted by its $y$-coordinates; each antenna is then assigned to the corresponding virtual element. The codebook uses an oversampled DFT with $Q\geq\sqrt M$, where $Q$ is the number of beam directions per dimension. The weakness of this baseline is that the phase of each beam depends only on the antenna's position in the hypothetical virtual array in this ordering, not on its actual location in the active array. For irregular active array selections, the assumed phase progression therefore differs from the true one, and the best codebook beam no longer matches the user's channel. The codebook can still be used, but with significant quantization errors. The next two codebooks address this.

\subsubsection{Physical-Grid DFT (PG-DFT) Codebook}
PG-DFT removes the position mismatch of VA-DFT in the simplest way.  Let $(n_y(i),n_z(i))$ denote the grid coordinates of the $i$th candidate antenna, with linear index $i=(n_z-1)N_y+n_y$. It builds a DFT dictionary for the full $N_y\times N_z$ candidate grid and keeps, for each active antenna, the entry corresponding to its physical grid location. Each antenna thus keeps the phase of its true grid position, even if its neighbors are inactive. Specifically, let
\begin{equation}
[\mathbf F_y]_{n,p}
=\frac{1}{\sqrt{N_y}}
e^{-j2\pi(n-1)(p-1)/Q_{\max}},
\end{equation}
where $\mathbf F_y\in\mathbb C^{N_y\times Q_{\max}}$ and define $\mathbf F_z\in\mathbb C^{N_z\times Q_{\max}}$ analogously, with $N_z$ rows. The full-grid dictionary is
\begin{equation}
\mathbf W_{\mathrm{full}}
=\mathbf F_z\otimes\mathbf F_y.
\end{equation}
Beam directions in this codebook are selected uniformly in the spatial-frequency domain of the full grid by taking DFT columns at intervals of $Q_{\max}/Q$, where $Q_{\max}$ is the DFT size of the full-grid dictionary and $Q_{\max}/Q$ is an integer. The directions are therefore fixed by the grid and do not depend on $\mathcal I$. For an active antenna index set $\mathcal I$, the PG-DFT dictionary is obtained by retaining the rows indexed by $\mathcal I$ and normalizing
each column to unit norm:
\begin{equation}
\mathbf W_{\mathrm{PG}}(\mathcal I)
=[\mathbf W_{\mathrm{full}}]_{\mathcal I,:}
\boldsymbol\Lambda(\mathcal I),
\end{equation}
where $\boldsymbol\Lambda(\mathcal I)$ is diagonal, with entries equal to the inverse norms of the corresponding restricted columns. Thus, each beam retains the phase of the antenna’s physical grid location. Because the direction set is inherited from the full grid, it is not tuned to the aperture of the selected antennas. When the active antennas span a smaller or sparser aperture than the full grid, the beamwidth and the resolution of the sampled directions differ from those the DFT assumes.

\subsubsection{Geometry-Aware (GA) Codebook}
Our main proposal is a GA codebook that constructs plane-wave steering vectors directly from the coordinates of the active antennas. \emph{Geometry-aware}
means that each beam is computed from the coordinates of the selected antennas, so it applies to any $\mathcal I$, regular or irregular, and its directions are chosen independently of the physical grid. Unlike PG-DFT, the direction set is not inherited from the full-grid dictionary. Concretely, let $\bar a_{y,m},\bar a_{z,m}$ be the in-plane coordinates of the active
antennas relative to their centroid. The GA codebook is
\begin{equation}\label{eq:ga_steering}
\mathbf w_{\mathrm{GA}}(u,v)=\tfrac{1}{\sqrt M}
\Big[e^{-j\frac{2\pi}{\lambda}(u\bar a_{y,m}+v\bar a_{z,m})}\Big]_{m=1}^{M},
\quad (u,v)\in\mathcal U^2 ,
\end{equation}
where the direction grid is $\mathcal U=\left\{-1+\frac{2(i-1)}{Q-1}\right\}_{i=1}^{Q}$, corresponding to $Q$ equally spaced points generated by $\texttt{linspace}(-1,1,Q)$. Each beam is the plane-wave response of the active antennas to the direction $(u,v)$, and $Q$ samples per direction coordinate give $Q^2$ beams.

\begin{figure}[t!]
        \centering         {\includegraphics[width=1\textwidth]{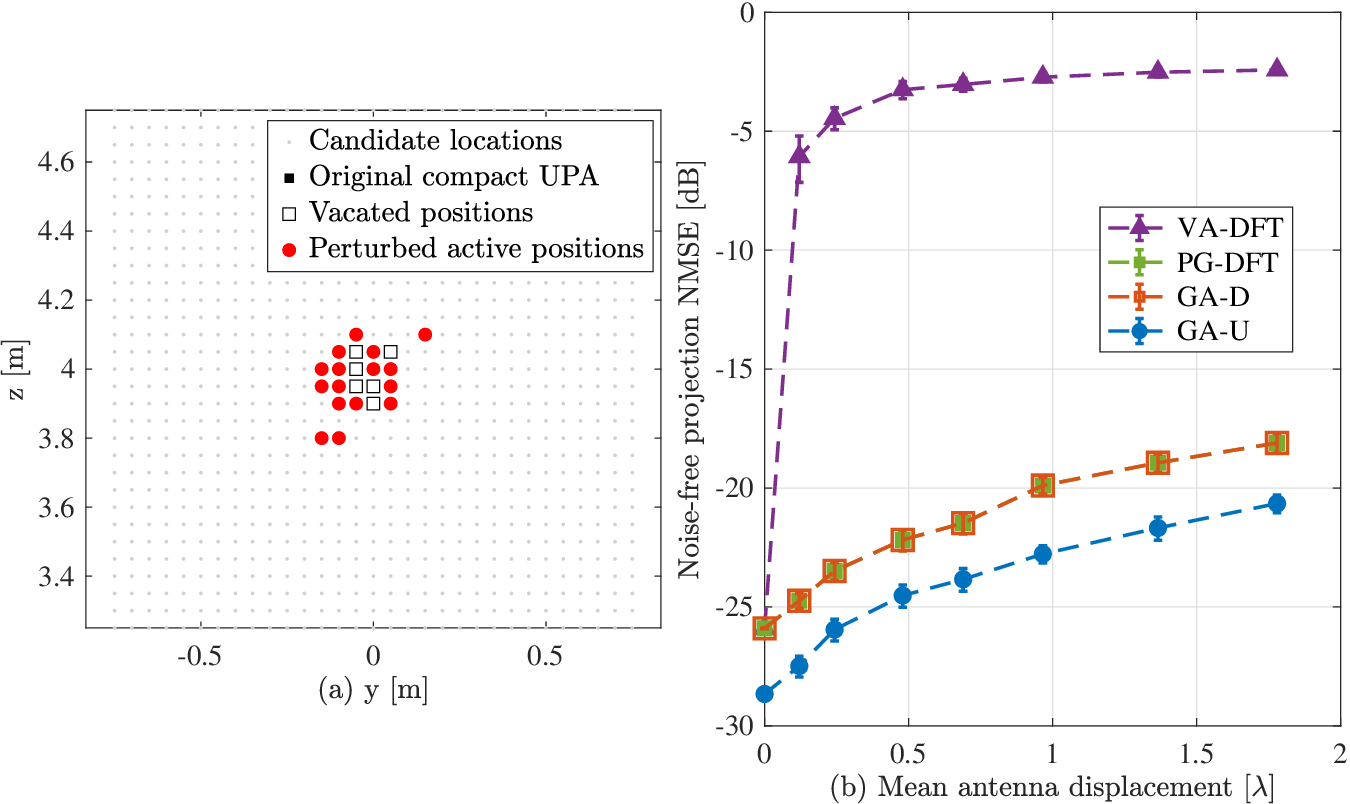}}
        \caption{Effect of antenna-position irregularity on channel projection NMSE. (a): original array (black squares) and an example with six displaced antennas (red circles); gray dots denote candidate positions. (b): noise-free NMSE versus normalized mean antenna displacement for the four codebooks.}
\label{figure_codebookirregularity}  \vspace{-2mm}
\end{figure} 

\subsection{Codebook Sensitivity to Antenna-Position Irregularity}\label{sec:codebook_irregularity}
The codebooks differ in how they treat the active antenna positions, so their projection accuracy should depend on how irregular the active array is. To quantify this effect, we measure the noise-free projection NMSE, defined as 
\begin{equation}
\mathrm{NMSE}
=\frac{\|\mathbf H(\mathcal I)-
\widehat{\mathbf H}_{\mathrm{cb}}(\mathcal I)\|_F^2}
{\|\mathbf H(\mathcal I)\|_F^2}.
\end{equation} 
for VA-DFT, PG-DFT, GA-D, and GA-U, where the projection is applied directly to the true LoS channel restricted to the active antennas ($\mathbf z_k=\mathbf h_k$). Starting from the compact $4\times4$ $\lambda/2$ UPA, $n\in\{0,1,2,4,6,8,12,16\}$ randomly chosen antennas are moved to random unoccupied candidate locations within $\pm4$ grid steps ($\pm2\lambda$) per axis, keeping $M=16$ fixed. Fig.~\ref{figure_codebookirregularity}(a) shows one realization: black squares are the original positions, red circles the final active positions, and gray dots the candidate locations. In Fig.~\ref{figure_codebookirregularity}(b), the NMSE is averaged over 100 user drops $K=10$ and 20 random perturbations per level, with $95\%$ confidence intervals across perturbations. The horizontal axis is the mean displacement over all $M$ active antennas, including unmoved ones, in units of $\lambda$.
We compare four codebooks: the virtual-array DFT (VA-DFT), which assigns the active antennas to virtual array elements according to their $z$- and $y$-coordinates; the physical-grid DFT (PG-DFT), obtained by extracting the active rows from the full-grid DFT dictionary; GA-DFT-aligned (GA-D), which samples the directions on the uniform endpoint-excluded grid $u_q=-1+2q/Q$, $q=0,\ldots,Q-1$; and GA-uniform (GA-U), which uses $Q=64$ equally spaced directions including both endpoints of $[-1,1]$ \cite{oversampling}.

For the regular array, PG-DFT and VA coincide ($\approx-26$~dB) since both use the same dictionary, while GA is about $2.7$~dB lower owing to its direction sampling. As displacement grows, the VA codebook degrades abruptly ($\approx-6$~dB at $0.1\lambda$) and saturates near $-2.5$~dB, because the coordinate-sorting assignment maps displaced antennas to wrong virtual positions and no codeword follows the resulting phase pattern. PG-DFT and GA degrade gradually, reaching about $-18$~dB and $-21$~dB at $1.8\lambda$, with GA consistently $2$-$3$~dB better. The results indicate that the choice of direction sampling substantially affects projection accuracy, while the geometry-aware formulation maintains more gradual degradation than the virtual-array baseline. The remaining projection error reflects the inherent limitations of using finite-sized dictionaries in representing the simulated channel, including spherical-wave effects (the codebooks assume plane waves) and the mismatch between the continuous user directions and the finite direction grid of the codebook. Because the experiment uses noise-free channel coefficients, the observed errors isolate codebook projection mismatch from observation-noise effects.

This motivates the use of geometry-aware codebooks when the active array is irregular, as in the antenna-selection problem studied next.

\subsection{Downlink Transmission and Performance Metric}\label{sec:performance_metric}
We assume the BS employs regularized zero-forcing (RZF) precoding with
\begin{equation}
\mathbf V_{\rm RZF}=\hat{\mathbf H}^H(\hat{\mathbf H}\hat{\mathbf H}^H+\alpha\mathbf I_K)^{-1}\in\mathbb C^{M\times K}
\end{equation}
where $\alpha=\frac{K\sigma_n^2}{P}$ is the regularization parameter and $P$ represents the total BS transmit power.
The precoder is based on the estimated channel $\hat{\mathbf H}\in\mathbb C^{K\times M}$, which can be either $\mathbf Z(\mathcal I)^T$ or $\widehat{\mathbf H}_{\mathrm{cb}}(\mathcal I)$ depending on whether codebook projection is used.
To satisfy the transmit power constraint while allocating equal power to each user stream, every precoding vector is first normalized to unit norm: $\tilde{\mathbf{v}}_k=\frac{\mathbf{v}_k}{{\| \mathbf{v}_k \|}_2}, k=1,2,\ldots,K$, where $\mathbf v_k$ denotes the $k$-th column of $\mathbf V_{\rm RZF}$. The normalized downlink precoder is then $\mathbf{V}=\sqrt{\frac{P}{K}}[\tilde{\mathbf{v}}_1,\cdots,\tilde{\mathbf{v}}_K]$, which allocates equal transmit power $P/K$ to every user.

We assume that the receiving user has perfect CSI during reception, so the true channel $\mathbf H=\mathbf H(\mathcal I)\in\mathbb C^{K\times M}$ is used when quantifying the achievable performance. The effective channel with RZF precoding is $\mathbf H_{\rm eff}=\mathbf H\mathbf V\in\mathbb C^{K\times K}$,
where $[\mathbf H_{\rm eff}]_{k,i}$ is the gain from $i$th stream to $k$th user. 

Thus, the signal-to-interference-and-noise ratio (SINR) of $k$-th user is 
\begin{equation}\label{sinr}
\mathrm{SINR}_k=\frac{
|[\mathbf{H}_{\mathrm{eff}}]_{k,k}|^2
}{\sum_{i\neq k}|[\mathbf{H}_{\mathrm{eff}}]_{k,i}|^2+
\sigma_n^2}.
\end{equation}

To evaluate the codebooks and the scanning strategies on a common baseline, we consider a statistical user model rather than a single geometry, so that the results do not depend on one particular user placement. Each user position $\mathbf b_k$ is drawn
independently and uniformly over the area of an annular sector $\mathcal D=\{(r,\phi): r_{\rm in}\le r\le r_{\rm out},\ |\phi|\le\pi/3\}$ in the horizontal plane in front of the BS, so $\mathbf B=[\mathbf b_1,\ldots,\mathbf b_K]$ has a distribution $p(\mathbf B)=\prod_{k=1}^K p(\mathbf b_k)$. Let $\tilde{\mathbf N}$ denote the pilot-noise realization. For a given $(\mathbf B,\tilde{\mathbf N})$, the channel $\mathbf H_{\rm grid}$ is determined by $\mathbf B$; the estimate, the selected configuration $\mathcal I$ and the precoder follow from $(\mathbf B,\tilde{\mathbf N})$; and the sum rate for this channel realization is
\begin{equation}
R(\mathbf B,\tilde{\mathbf N})=\sum_{k=1}^{K}\log_2\!\big(1+\mathrm{SINR}_k\big).
\end{equation}
The performance metric is the statistical average sum rate
\begin{equation}\label{average_sumrate}
\bar R_{\rm sum}=\mathbb E_{\mathbf B,\tilde{\mathbf N}}\big[R(\mathbf B,\tilde{\mathbf N})\big].
\end{equation}
This expectation has no closed form, so we approximate it by Monte Carlo simulation. We draw $N_{\rm MC}$ i.i.d. user drops $\mathbf B_1,\ldots,\mathbf B_{N_{\rm MC}}\sim p(\mathbf B)$ and, for each drop, $R_{\rm avg}$ i.i.d. pilot-noise realizations $\tilde{\mathbf N}_{n,1},\ldots,\tilde{\mathbf N}_{n,R_{\rm avg}}$, and compute
\begin{equation}\label{eq:mc_estimate}
\bar R_{\rm sum}\approx\frac{1}{N_{\rm MC}R_{\rm avg}}
\sum_{n=1}^{N_{\rm MC}}\sum_{r=1}^{R_{\rm avg}}R(\mathbf B_n,\tilde{\mathbf N}_{n,r}).
\end{equation}
By the law of large numbers, \eqref{eq:mc_estimate} converges to \eqref{average_sumrate} as $N_{\rm MC}\to\infty$. The values of $N_{\rm MC}$ and $R_{\rm avg}$ are chosen large enough that the sample average is a close approximation of the statistical average.

\begin{figure}[t!]
        \centering         {\includegraphics[width=0.95\textwidth]{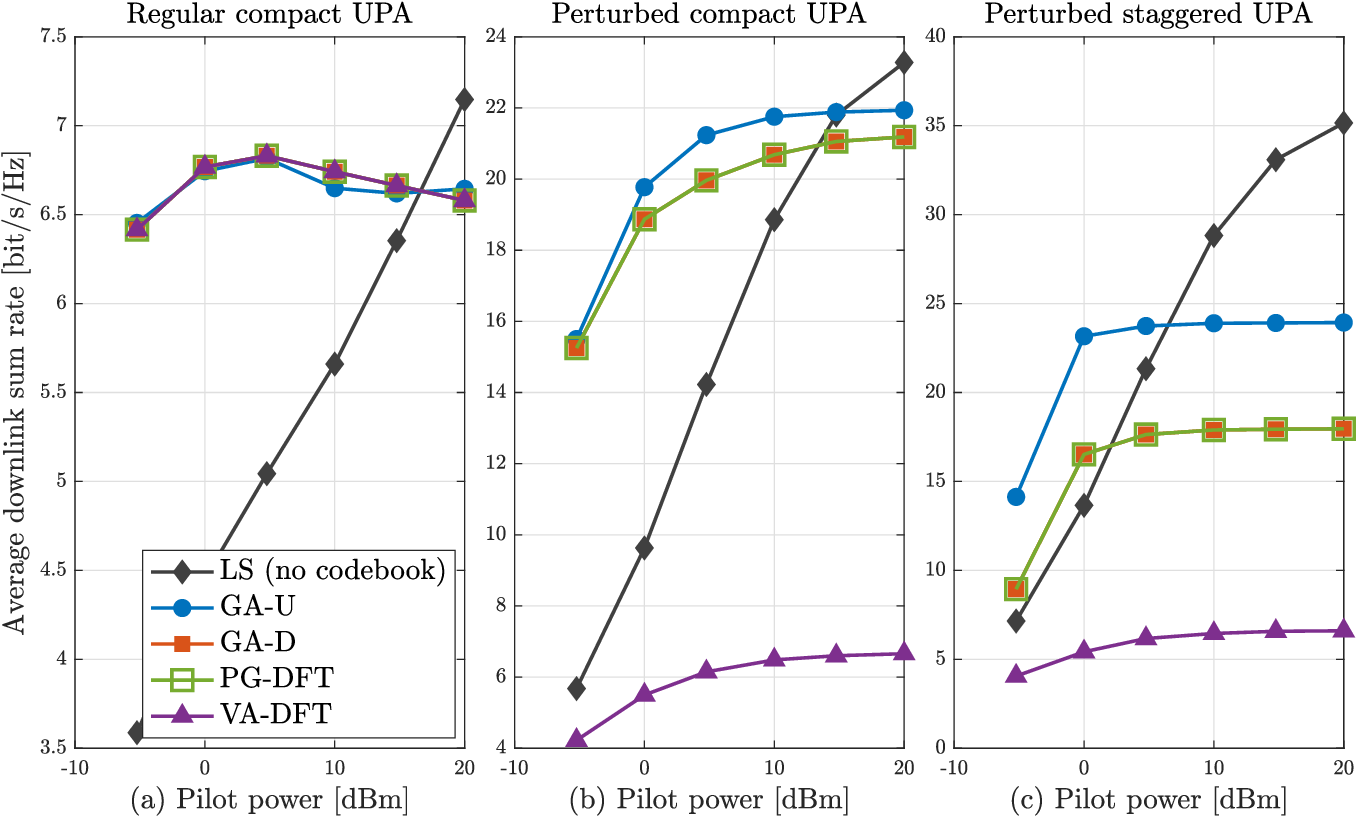}}
        \caption{Average sum rate versus pilot power $P_{\mathrm{ul}}$ at fixed downlink power $P=20$\,dBm, for: (a)~regular compact UPA; (b)~severely perturbed compact UPA; and (c)~severely perturbed staggered UPA.}
\label{figure_snr_plot}  \vspace{-2mm}
\end{figure}

To examine the coupling between pilot power, channel acquisition, and downlink performance, Fig.~\ref{figure_snr_plot} compares LS channel estimation without codebook projection with codebook-based estimation using GA-U, GA-D, PG-DFT, and VA-DFT. We consider three scenarios: (i) a regular compact $4\times4$ UPA with $\lambda/2$ spacing; (ii) the same compact UPA with all 16 antennas relocated to random candidate locations; and (iii) a staggered UPA, a fixed non-uniform planar array with $4\times4$ elements, horizontal spacing $4\lambda$ and vertical spacing $5\lambda$, and a progressive horizontal shift of $\lambda$ between successive rows, with all 16 antennas relocated as in (ii). All positions lie on the $\lambda/2$ candidate grid. Unless otherwise noted, the results use the parameters in Table~\ref{tab:sim_parameters}. For the perturbed configurations, results are averaged over 20 random perturbations and 100 user drops. The same user drops and channel realizations are used across the schemes being compared to ensure a fair comparison and reduce variability in their relative performance.

\begin{table}[t!]
\centering
\caption{Default Simulation Parameters}
\small
\setlength{\tabcolsep}{3pt}
\begin{tabular}{l|l|l|l}
\hline
\textbf{Parameter} & \textbf{Value} & \textbf{Parameter} & \textbf{Value}\\
\hline
BS height & $4.25$ m & BS transmit power & $20$ dBm\\
\hline
Carrier frequency & $3$ GHz & Bandwidth & $100$ MHz\\
\hline
Number of users & $10$ & Active antennas & $16$\\
\hline
User height & $1.5$ m & User drops & 100\\
\hline
Noise figure & $10$ dB & Noise power & $-84$ dBm\\
\hline
\end{tabular}
\label{tab:sim_parameters}
\end{table}
The results reveal a trade-off between pilot-noise suppression and channel-representation error. The LS estimate is unbiased, with error variance $\sigma_n^2/(\tau_p P_{\mathrm{ul}})$, which decreases as the uplink pilot power $P_{\mathrm{ul}}$ increases. Codebook-based projection suppresses the pilot noise but introduces a representation error due to the finite direction grid and the single-beam approximation. Consequently, GA-U achieves higher sum rates at low pilot powers but eventually saturates, whereas the LS rate continues to improve. At $P_{\mathrm{ul}}=0$~dBm, GA-U achieves approximately twice the LS rate on the perturbed compact array and about $70\%$ higher rate on the perturbed staggered array. LS overtakes GA-U at approximately $15$~dBm for the compact array and $6$-$7$~dBm for the staggered UPA. PG-DFT exhibits similar saturation but achieves a lower rate on the staggered UPA, where row extraction can reduce the effective spatial resolution of the dictionary. VA-DFT performs worse than LS on irregular arrays because its representation error outweighs the noise reduction it provides. On the regular compact array, the codebooks coincide under the considered array and codebook construction.
This saturation is not a fundamental limit of codebook projection. It reflects the fixed codebook size ($Q=64$) and the single-beam model: for a given $Q$, the representation error sets a floor, and projection helps only while the pilot-noise error exceeds this floor. The crossover power therefore depends on the codebook and on the array. A denser codebook lowers the off-grid part of the floor and moves the crossover to higher pilot power, at the cost of a larger search space. The remaining part, namely the spherical-wave mismatch and multipath beyond one beam, cannot be removed by increasing $Q$. 
To clarify the operating regime, note that with $f_c=3$ GHz and users at 200-300 m, the free-space path loss is about 88-92 dB. For $\tau_p=K=10$ and $-84$ dBm noise power, the per-antenna LS estimation SNR $\tau_pP_{\rm ul}\beta/\sigma_n^2$ at a user at 250 m is about 4 dB for $P_{\rm ul}=0$ dBm and 9 dB for $P_{\rm ul}\approx4.8$ dBm (3 mW), the value used in the remaining results; across the 200-300 m sector, the range $P_{\rm ul}\in[0,5]$ dBm corresponds to roughly 3-11 dB. This is a low-to-moderate SNR regime, representative of power-limited pilots or, equivalently, of a higher-power user experiencing additional shadowing or blockage loss. For reference, the downlink SNR per user with perfect CSI and no interference is about 16 dB (20 dBm total power, equal allocation, 16-antenna array gain). Since the pilot power available to users is typically limited, we focus on $P_{\rm ul}\in[0,5]$~dBm in the remainder of the paper, where GA outperforms LS in all three scenarios.

For the considered uniform \(\lambda/2\) candidate lattice, GA-D and PG-DFT employ equivalent spatial sampling under the adopted direction-grid, coordinate-ordering, and normalization conventions. Their dictionaries are therefore identical up to column permutation, resulting in overlapping performance curves. The comparison between GA-U and GA-D highlights the effect of direction-grid selection, while the comparison between GA-D and PG-DFT confirms the equivalence of the geometry-aware and physical-grid DFT constructions under these conditions.

\section{Discrete Antenna Geometry Optimization}
\label{sec:optimization}
The positions of the active antennas determine both the propagation channel and the accuracy of its estimate, so antenna placement and downlink performance are tightly coupled. This section formulates the selection of the $M$ active antennas as an optimization problem, introduces a score that the BS can evaluate based on its available CSI, and describes the search algorithm that maximizes it. The selection problem and the score defined here are reused in the following sections, where they serve as the basis for the proposed geometry optimization and its comparison with the benchmark schemes.

\subsection{Antenna Selection Problem}
For a given user drop, we select $M$ antennas from the scanned set $\mathcal S_{\rm scan}$ to maximize the sum rate of Sec.~\ref{sec:performance_metric}:
\begin{equation}
\max_{\mathcal I\subseteq\mathcal S_{\rm scan},\,|\mathcal I|=M} R(\mathcal I).
\end{equation}
The rate $R(\mathcal I)$ depends on the true channel, whereas the BS only has pilot-based channel estimates. We therefore compare candidate sets using a score computed entirely from these estimates. 
\begin{figure*}[t!]
    \centering
    \subfloat[Staggered UPA.]{
        \includegraphics[width=0.26\textwidth]
        {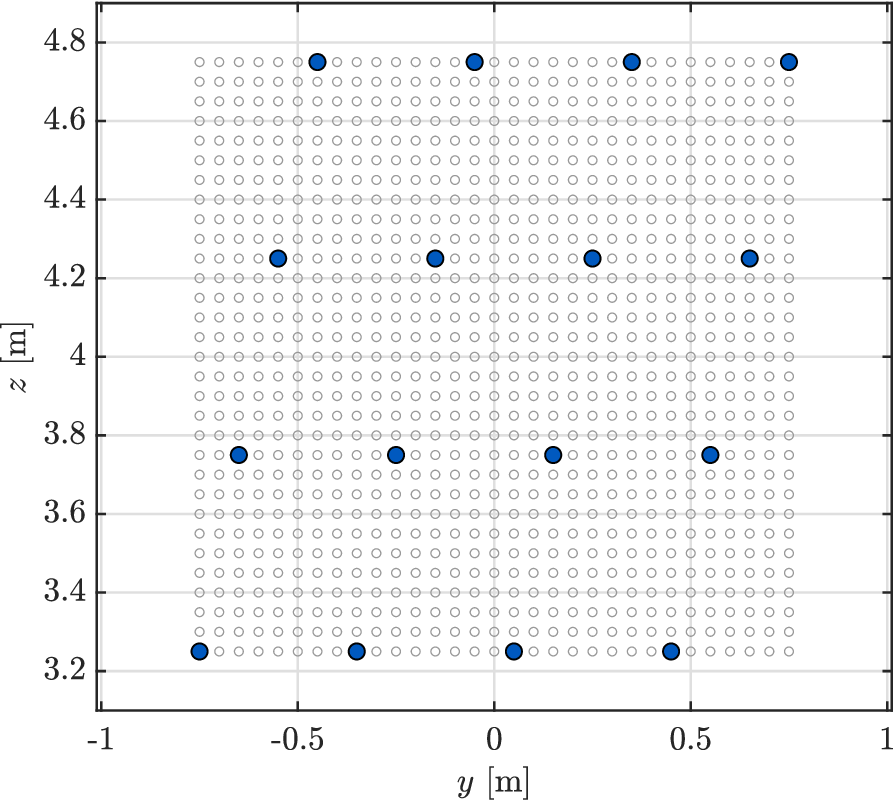}}
    \hfill
    \subfloat[Sparse UPA.]{
        \includegraphics[width=0.26\textwidth]
        {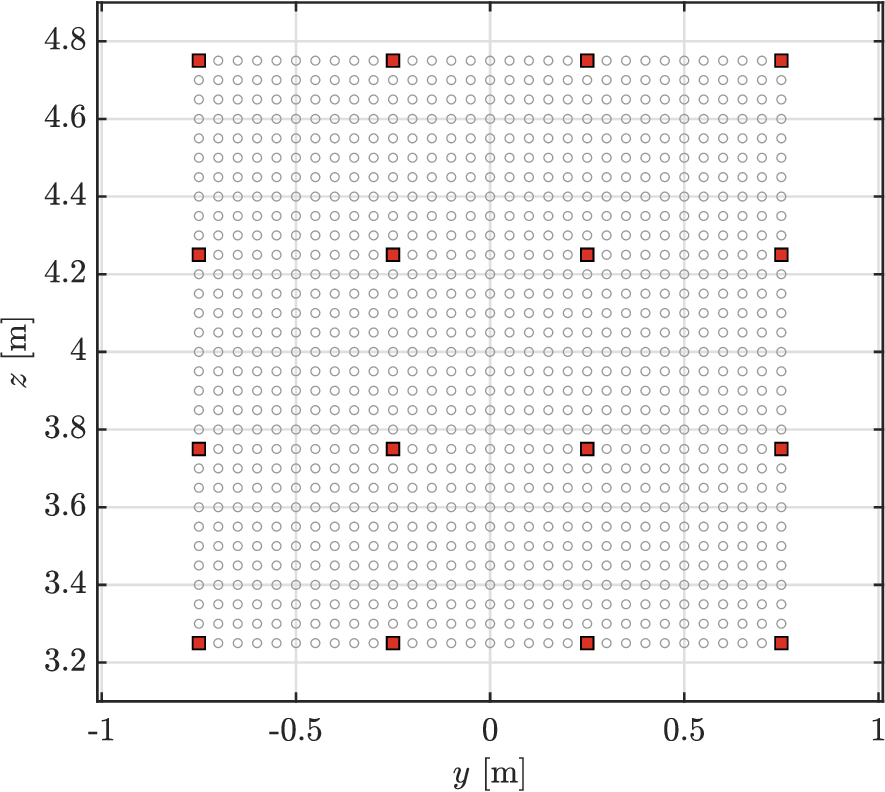}}
    \hfill
    \subfloat[Compact UPA.]{
        \includegraphics[width=0.26\textwidth]
        {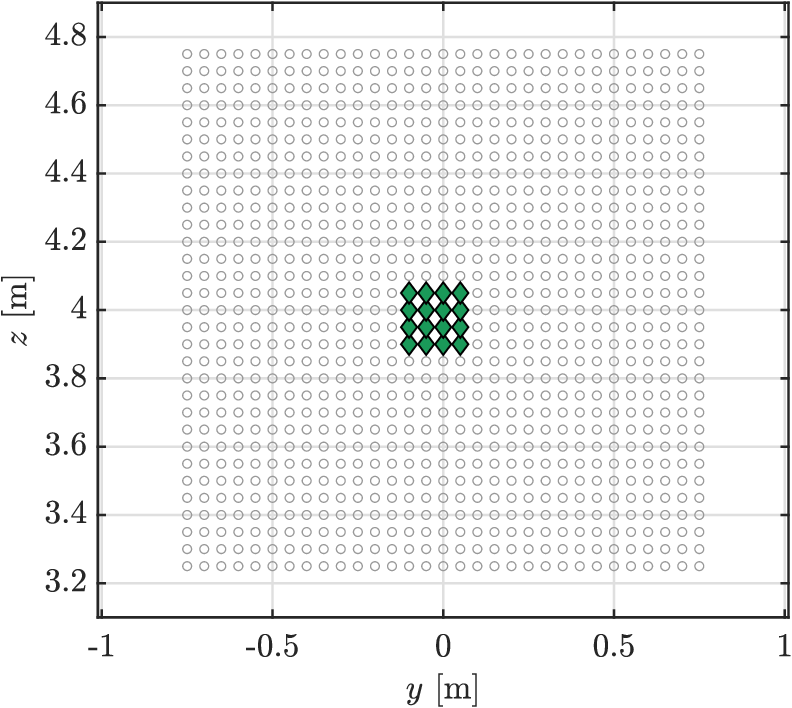}}
    \hfill
    \subfloat[User distribution.]{
        \includegraphics[width=0.15\textwidth]
        {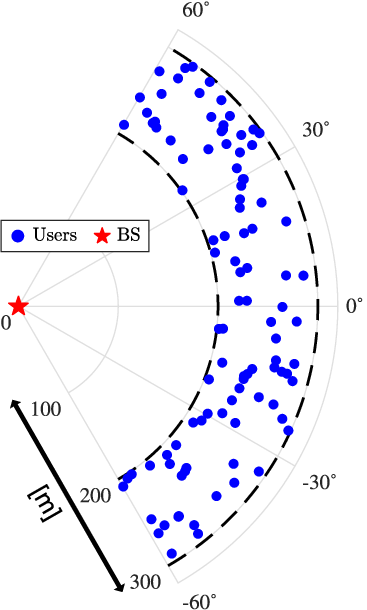}}
    \caption{Benchmark antenna geometries and user distribution.}
    \label{Benchmark schemes}
\end{figure*}

\subsection{Train/Test Score}
To compare antenna sets without the true channel, the BS measures the scanned antennas twice with independent pilot noise, giving a \emph{design} estimate $\widehat{\mathbf H}^{\rm tr}$ and a \emph{test} estimate $\widehat{\mathbf H}^{\rm te}$. For a candidate set $\mathcal I$, the BS computes the RZF precoder $\mathbf V^{\rm tr}(\mathcal I)$ from the rows of $\widehat{\mathbf H}^{\rm tr}$ indexed by $\mathcal I$ and scores it on the corresponding rows of $\widehat{\mathbf H}^{\rm te}$:

\begin{equation}
J(\mathcal I)=\sum_{k=1}^{K}\log_2\!\big(1+\mathrm{SINR}^{\rm te}_k(\mathcal I)\big),
\end{equation}
where $\mathrm{SINR}^{\rm te}_k$ is computed as in \eqref{sinr} with the test channel in place of the true channel. Since the test noise is independent of the training noise, a precoder that overfits the training noise gains nothing in $J$, so $J$ is a reliable proxy for $R(\mathcal I)$.

\subsection{Greedy Search}
Exhaustive search over all $\binom{M_{\rm grid}}{M}$ sets is prohibitive in the practically-sized systems considered in this paper. Algorithm~\ref{alg:greedy_cb} instead updates one antenna at a time, first over a coarse grid and then locally around the best coarse candidate, while keeping the others fixed. Its cost is $O\big(MN_{\rm pass}(N_c+N_l)\big)$ evaluations of $J$, where $N_c$ and $N_l$ are the numbers of coarse and local candidates per antenna and $N_{\rm pass}$ is the number of passes.

\begin{algorithm}[t!]
\small
\caption{Hierarchical Greedy Antenna Selection}
\label{alg:greedy_cb}
\begin{algorithmic}[1]
\Require $\mathcal S_{\rm scan}$, warm start $\mathcal I^{(0)}$, score $J(\cdot)$, passes $N_{\rm pass}$
\Ensure Selected set $\mathcal I^{\rm opt}$, estimate $\widehat{\mathbf H}$
\State Initialize $\mathcal I$ from $\mathcal I^{(0)}\cap\mathcal S_{\rm scan}$, padded to $M$ antennas
\For{$p=1,\ldots,N_{\rm pass}$}
  \For{$m=1,\ldots,M$}
    \State \textbf{Coarse:} try replacing the $m$-th antenna with each coarse-grid candidate; keep the best $c^\star$
    \State \textbf{Local:} try every candidate in a window around $c^\star$
    \State \textbf{Accept} the best replacement if it increases $J$
  \EndFor
  \If{$\mathcal I$ did not change} \textbf{break} \EndIf
\EndFor
\State $\widehat{\mathbf H}\gets$ codebook projection of the train rows at $\mathcal I$ (none for LS)
\State \Return $\mathcal I,\widehat{\mathbf H}$
\end{algorithmic}
\end{algorithm}

\section{Codebook Performance Under Full Spatial Scanning}\label{sec:fullscanning}

In this section, the performance of the optimized MA system is evaluated against three fixed array geometries, each comprising $M=16$ antennas. The MA selects $M=16$ antennas from a $31\times31$ candidate grid ($M_{\rm grid}=961$) with $\lambda/2$ spacing, i.e., a $15\lambda\times15\lambda$ ($1.5\,\text{m}\times1.5\,\text{m}$) aperture. The benchmark geometries are illustrated in Fig.~\ref{Benchmark schemes}. To focus on the ultimately achievable performance, we consider full spatial scanning in this section, while partial scanning will be considered later.
The considered antenna geometries are as follows:

\begin{itemize}
  \item \textbf{MA}: A movable antenna array whose element positions are selected from $A_{grid}$ using
    Algorithm~\ref{alg:greedy_cb}.
  \item \textbf{Staggered UPA}: A non-uniform planar array with horizontal spacing of $4\lambda$ with alternate rows offset and with vertical spacing of $5\lambda$.
  \item \textbf{Sparse UPA}: A uniform planar array with both horizontal and vertical spacing of $5\lambda$.
  \item \textbf{Compact UPA}: A uniform planar array with both horizontal and vertical spacing of $\lambda/2$.
\end{itemize}

We first evaluate the achievable sum rate under different channel estimation schemes, which differ only in how $\mathbf z_k$ is processed:
\begin{itemize}
  \item \textbf{LS}: The unprojected least-squares estimate $\mathbf z_k$.
  \item \textbf{VA-DFT}: Projection onto the virtual-array DFT codebook.
  \item \textbf{PG-DFT}: Projection onto the physical-grid DFT codebook.
  \item \textbf{GA-U}: Projection onto the geometry-aware codebook.
\end{itemize}

To isolate the loss caused by imperfect CSI acquisition, we also include an \textbf{MA-Perfect CSI} benchmark. It uses the same optimized MA geometry from Algorithm~\ref{alg:greedy_cb} but precodes with the true channel $\mathbf H$ instead of the estimate $\widehat{\mathbf H}$. The gap between this benchmark and the practical MA results therefore quantifies the rate loss due to CSI acquisition under identical geometry and simulation settings.
\subsubsection{Average Sum Rate Across Array Geometries}
\begin{figure}[t!]
        \centering         {\includegraphics[width=1\textwidth]{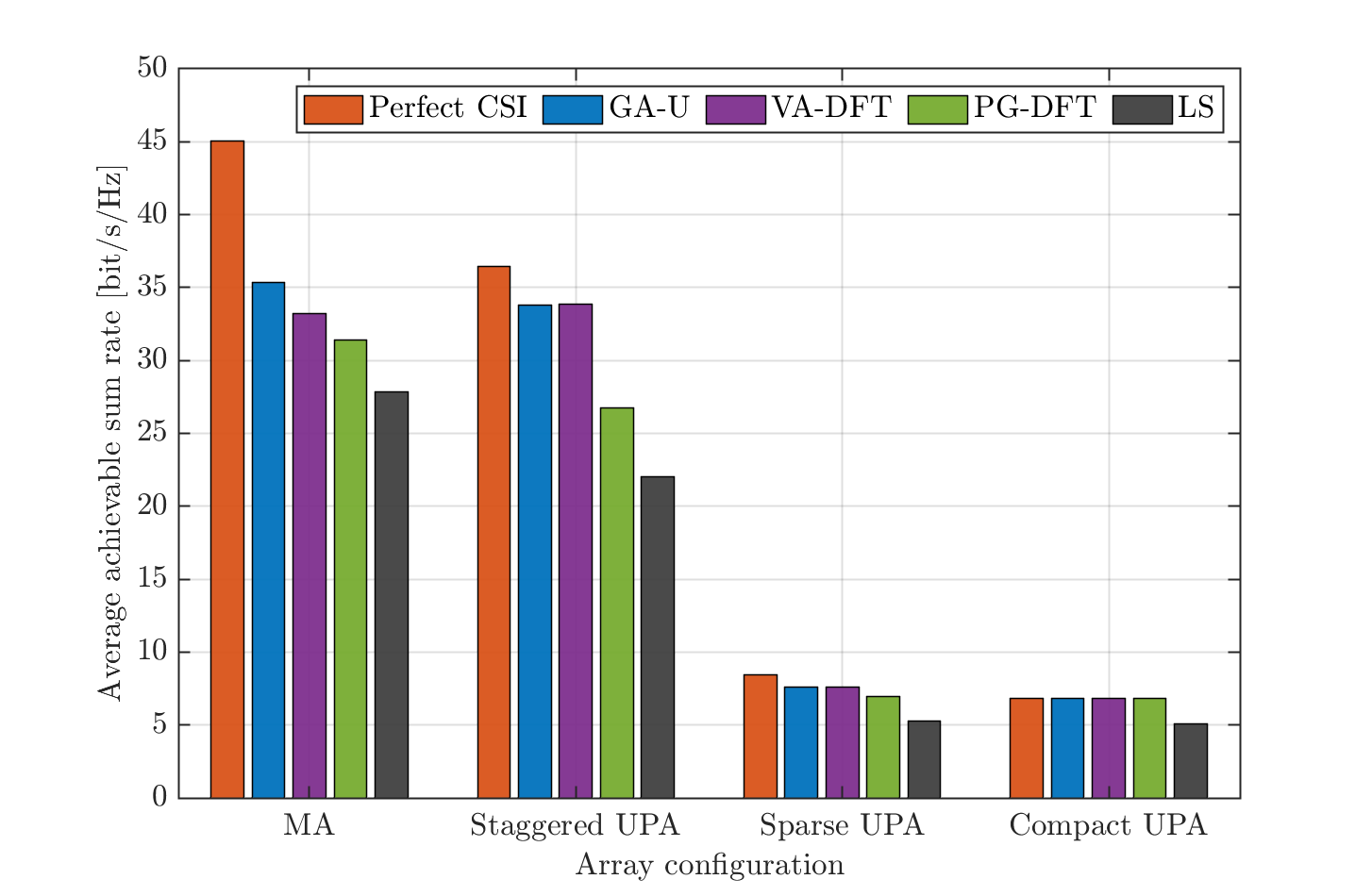}}
        \caption{The average achievable sum rate for all array configurations for the four acquisition methods. }
\label{figure_average}  \vspace{-2mm}
\end{figure}

Fig.~\ref{figure_average} reports the average sum rate of all array geometries with the four channel estimation schemes and, as a reference, perfect CSI. Three observations follow. First, with perfect CSI, the optimized MA achieves $45.1$~bit/s/Hz, which is $23.6\%$ above the Staggered UPA and about $5.3\times$ the Sparse UPA, confirming the massive potential benefit of position optimization. Second, the practical gain depends strongly on the codebook. With GA-U, the MA attains $35.4$~bit/s/Hz ($78.5\%$ of perfect CSI), a gain of $6.5\%$ over VA-DFT, $12.6\%$ over PG-DFT and $27.0\%$ over LS. With VA-DFT, the MA is even slightly below the Staggered UPA, so a geometry-agnostic codebook can cancel the advantage of the optimized array, whereas GA-U retains a $4.5\%$ lead over the best Staggered result. Third, for the regular arrays, the codebook choice matters less: on the Staggered UPA, GA-U and VA-DFT are essentially equal ($92.6\%$ of perfect CSI) and both clearly exceed PG-DFT and LS. The Sparse UPA behaves similarly. The Compact UPA is insensitive to the codebook ($\approx6.8$ [bit/s/Hz] for GA-U, VA-DFT, PG-DFT and Perfect CSI), because strong spatial correlation at $\lambda/2$ spacing limits the rate regardless of CSI quality; only LS falls behind, as it receives no denoising. Overall, projecting onto any codebook improves on LS for all geometries, and the geometry-aware formulation is the only scheme that keeps the MA ahead of the structured arrays.
\subsubsection{Sum-Rate Distribution for MA and Staggered UPA} 

\begin{figure}[t!]
        \centering         {\includegraphics[width=1\textwidth]{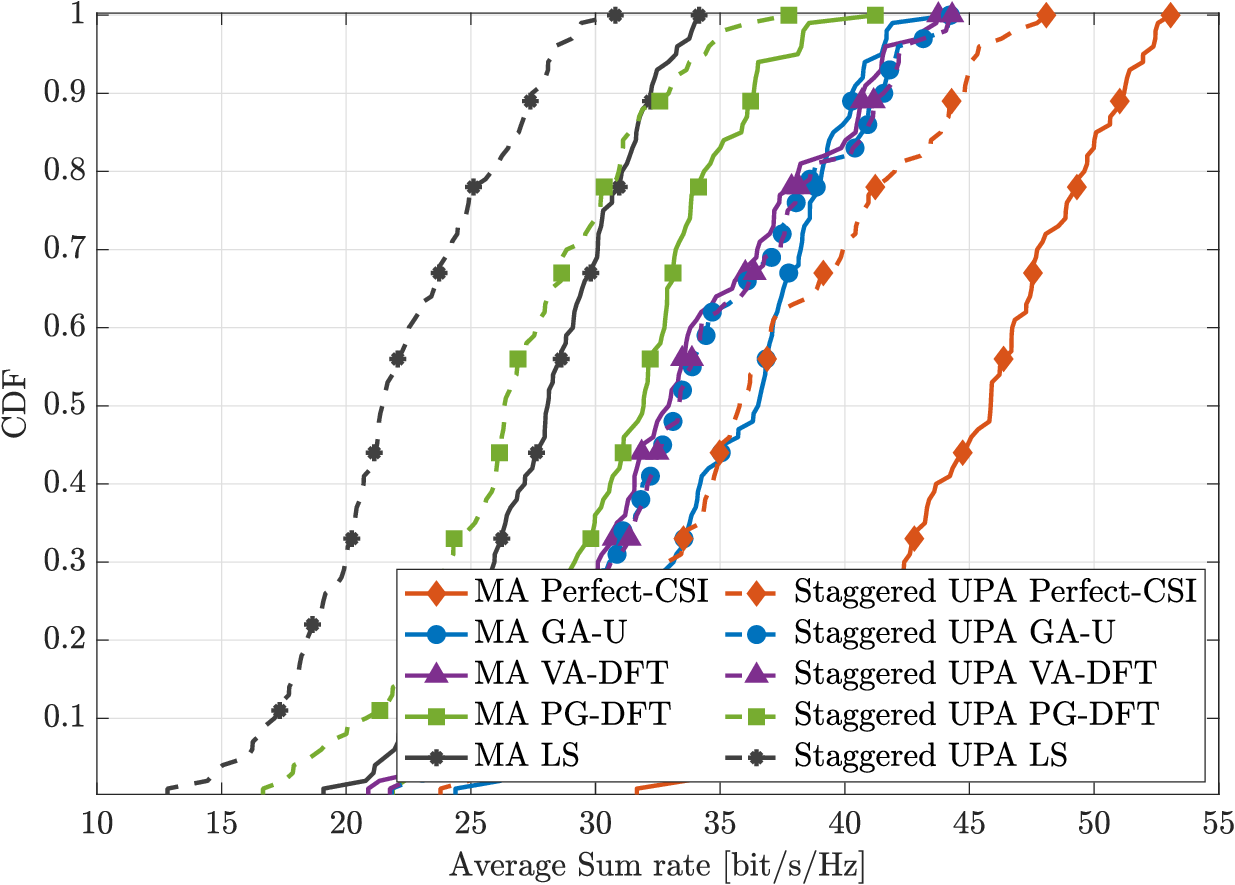}}
        \caption{CDF of the average achievable sum rate for all array configurations for the four channel estimation methods. }
\label{figure_cdf}  \vspace{-2mm}
\end{figure} 

Fig.~\ref{figure_cdf} shows CDFs of the average sum-rate of the optimized MA and the Staggered UPA under LS, VA-DFT, PG-DFT, GA-U, and perfect CSI, where the randomness is $100$ user drops shared by all schemes. For the MA, the GA-U curve lies to the right of the VA-DFT, PG-DFT and LS curves over most of the range. The median sum rate is $36.5$~bit/s/Hz with GA-U, a gain of $10.9\%$, $14.5\%$ and $30.1\%$ over VA-DFT, PG-DFT and LS, respectively. GA-U also outperforms VA-DFT in $74\%$ of the realizations and PG-DFT in $97\%$. GA-U narrows the distribution relative to VA-DFT (standard deviation $4.3$ versus $5.5$~bit/s/Hz). For the Staggered UPA, the GA-U and VA-DFT curves are similar, whereas PG-DFT and LS are clearly shifted to the left, with the gap to perfect CSI only $2.4$~bit/s/Hz. Moreover, with GA-U the MA exceeds the Staggered UPA at the median by $9.6\%$, while with VA-DFT this advantage vanishes. These trends agree with the averages in Fig.~\ref{figure_average}: geometry-dependent projection matters most for the irregular MA, whereas the structured array is already well represented by the virtual-array DFT. 

The full-scan MA does not exceed the staggered UPA with VA-DFT, as shown in Fig.~\ref{figure_average}, so VA-DFT cannot reveal the benefit of position optimization, whereas GA-U retains it. Also, GA-U outperforms PG-DFT on both arrays (MA and Staggered UPA), so PG-DFT also adds nothing to the scanning question. Since GA-U, or simply GA, is the best-performing practical scheme, the remainder of the paper focuses on the optimized MA with GA, and uses the MA with LS, which has no codebook projection, as the unprojected reference.

\section{Scanning-Aware Antenna Optimization}\label{sec:scanning}
The results so far considered the full candidate grid with the scanning set $\mathcal S_{\mathrm{scan}}=\mathcal A_{\mathrm{grid}}$, which will incur a significant training overhead in practice. Hence, the interesting question is whether it is possible to reduce the scanning region rather than probing all candidate grid positions, and, if so, whether the scanning method matters. Thus the scanning set $\mathcal S_{\mathrm{scan}}\subseteq\mathcal A_{\mathrm{grid}}$ can be
produced directly by a scanning policy $\pi$,
\begin{equation}
\mathcal S_{\mathrm{scan}}=\pi(\rho_{\mathrm{scan}}),\qquad
|\mathcal S_{\rm scan}|=\lceil\rho_{\rm scan}M_{\rm grid}\rceil.
\end{equation}
Because the hardware can only acquire pilots for $M$ locations at a time, the
locations in $\mathcal S_{\mathrm{scan}}$ are subsequently partitioned into pilot
batches $\mathcal T_1,\ldots,\mathcal T_{n_b}$, each of size at most $M$, with
$\mathcal S_{\mathrm{scan}}=\bigcup_{b=1}^{n_b}\mathcal T_b$. To analyze to what extent the scanning procedure matters, we consider two fixed-fraction scanning policies and one adaptive policy:

\begin{itemize}
    \item \textbf{Uniform-lattice scanning ($\pi_{\mathrm{unif}}$):}
    The scanning locations are selected according to a fixed, nested ordering of the candidate grid. Scanning begins with a coarse square lattice and progressively adds locations between existing lattice points, first using a quincunx-like pattern (four points around a central point) and then completing a finer square lattice. This refinement process provides increasingly dense spatial coverage while ensuring that the scanned sets are nested as $\rho_{\mathrm{scan}}$ increases. The policy is deterministic and does not adapt to the observed channel realization.

    \item \textbf{Random-individual scanning ($\pi_{\mathrm{rand}}$):}
    A prescribed number of distinct candidate locations is selected uniformly at random without replacement. Unlike uniform-lattice scanning, this policy does not guarantee spatial coverage, and the selected pattern varies across Monte Carlo repetitions.

    \item \textbf{Adaptive scanning ($\pi_{\mathrm{adapt}}$):} Unlike the two fixed-fraction policies, which determine the number of scanned locations in advance, adaptive scanning uses the acquired channel measurements to decide when further scanning is no longer worthwhile. Locations are measured in batches following a fixed exploration order. The first batch consists of the initial array $\mathcal I^{(0)}$, followed by its neighbors within a local window, and then the remaining locations. Let $\mathcal S_b=\bigcup_{j\le b}\mathcal T_j$ be the set scanned after batch $b$, and let
\begin{equation*}
E_b=\frac{1}{K|\mathcal S_b|}\big\|[\widehat{\mathbf H}_{\rm grid}]_{\mathcal S_b,:}\big\|_F^2,\qquad
\delta_b=\frac{|E_b-E_{b-1}|}{E_{b-1}}
\end{equation*}
be the average estimated channel energy and its relative change. Scanning stops at the first batch $b\ge n_{\min}$ for which $\delta_{b'}<\eta$ for the last $P$ consecutive batches ($P$ is the patience), or when $|\mathcal S_b|\ge\rho_{\max}M_{\rm grid}$. Hence, $\rho_{\rm eff}$ depends on the channel realization, and $\eta$ controls the trade-off between scanning cost and rate. The procedure is summarized in Algorithm~\ref{alg:adaptive_scan}.
\end{itemize}

\begin{figure}[t!]
    \centering
    \includegraphics[width=0.32\textwidth]{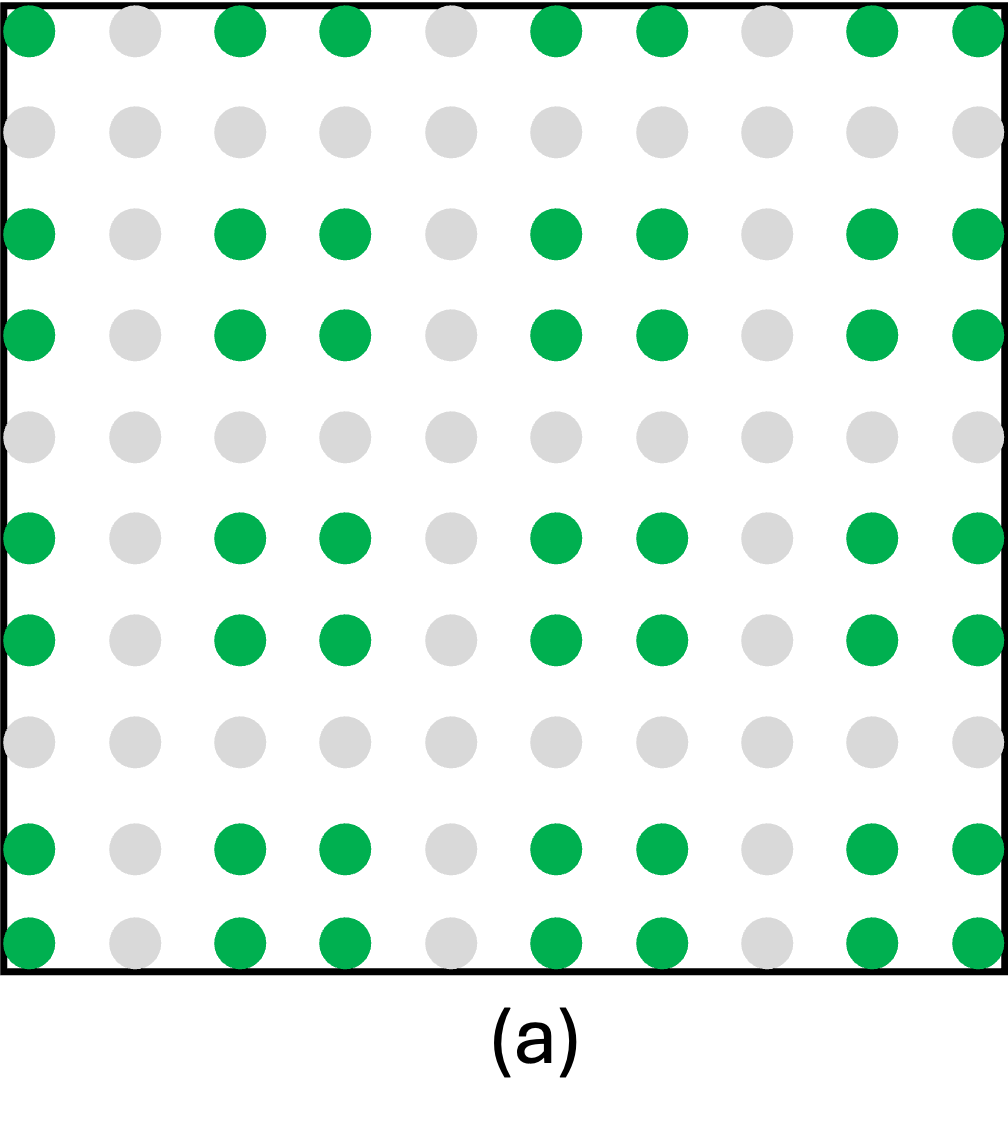}
    \includegraphics[width=0.32\textwidth]{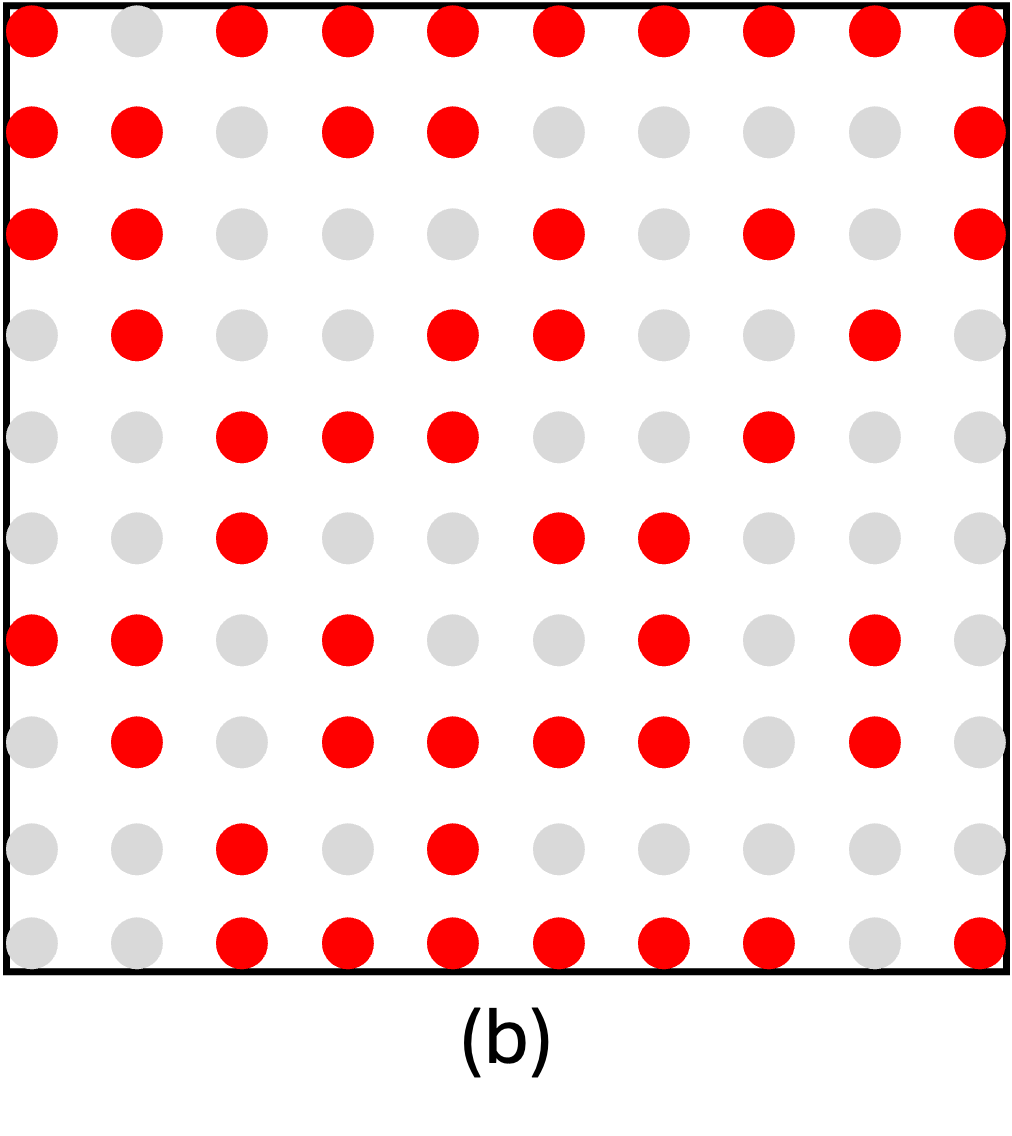}
    \caption{Illustration of the two scanning policies: uniform-lattice  $\pi_{\mathrm{unif}}$, and random-individual $\pi_{\mathrm{rand}}$. Colored markers denote scanned locations, gray markers denote unscanned locations. (a) $\pi_{\mathrm{unif}}$: a deterministic, evenly spaced sub-lattice covering the full aperture.(b) $\pi_{\mathrm{rand}}$: distinct locations sampled uniformly at random without replacement.}
    \label{fig:scan_policies}
\end{figure}

 Fig.~\ref{fig:scan_policies} depicts a schematic diagram of the two fixed policies on a reduced candidate grid at $\rho_{\mathrm{scan}}=0.5$. The uniform-lattice scanning in Fig.~\ref{fig:scan_policies}a gives deterministic, evenly distributed coverage across the aperture, while the random-individual scanning in Fig.~\ref{fig:scan_policies}b gives scattered coverage that differs across repetitions. For partial scanning, the scan set is produced by $\pi_{\rm unif}$, $\pi_{\rm rand}$, or Algorithm~\ref{alg:adaptive_scan}, and Algorithm~\ref{alg:greedy_cb} is then run on it, with the coarse and local candidates restricted to $\mathcal S_{\rm scan}$. If an antenna has no candidate in $\mathcal S_{\rm scan}$ within its window, it keeps its current position. The next section evaluates these policies.

\begin{algorithm}[t!]
\small
\caption{Adaptive Scanning $\pi_{\mathrm{adapt}}$}
\label{alg:adaptive_scan}
\begin{algorithmic}[1]
\Require Initial array $\mathcal I^{(0)}$, exploration order, threshold $\eta$, patience $P$, minimum batches $n_{\min}$, maximum fraction $\rho_{\max}$
\Ensure Scan set $\mathcal S_{\rm scan}$, train/test estimates on $\mathcal S_{\rm scan}$
\State $\mathcal S\gets\emptyset$, $c\gets0$
\For{$b=1,2,\ldots$}
  \State $\mathcal T_b\gets$ next $M$ unscanned locations in the exploration order
  \State Measure $\mathcal T_b$ twice (train, test); $\mathcal S\gets\mathcal S\cup\mathcal T_b$
  \State Compute $E_b$
\If{$b\ge2$}
  \State $\delta_b\gets|E_b-E_{b-1}|/E_{b-1}$
  \State \textbf{if} $\delta_b<\eta$ \textbf{then} $c\gets c+1$ \textbf{else} $c\gets0$
\EndIf
  \If{$(b\ge n_{\min}\ \textbf{and}\ c\ge P)$ \textbf{or} $|\mathcal S|\ge\rho_{\max}M_{\rm grid}$} \textbf{break} \EndIf
\EndFor
\State \Return $\mathcal S_{\rm scan}\gets\mathcal S$ \Comment{then run Algorithm~\ref{alg:greedy_cb} on $\mathcal S_{\rm scan}$}
\end{algorithmic}
\end{algorithm}

\begin{figure*}[t!]
    \centering
    \subfloat[Crossover between MA-GA and fixed scheme.]{ \includegraphics[width=0.45\textwidth]{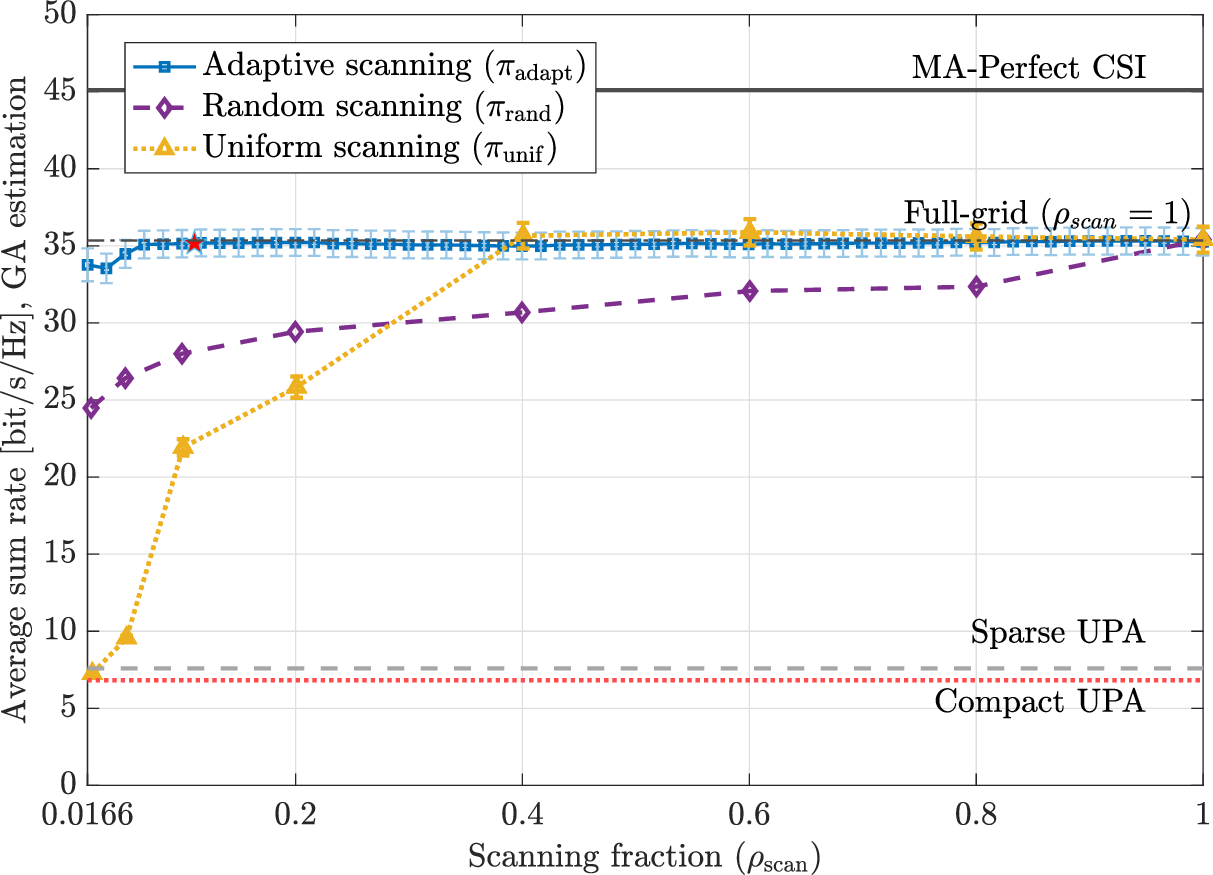}
        \label{GA_scan}}
    \subfloat[Crossover between MA-LS and fixed scheme.]{ \includegraphics[width=0.45\textwidth]{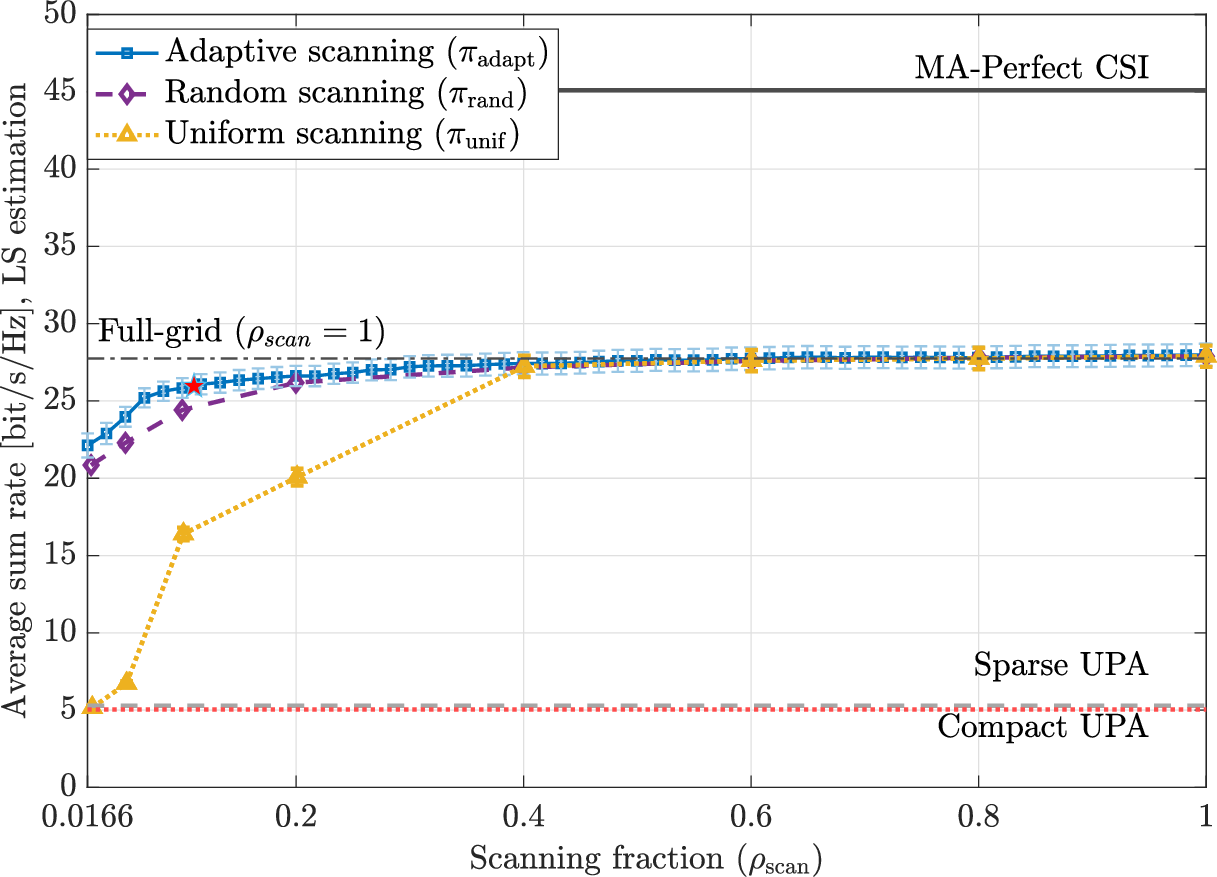}
        \label{LS_scan}}    
        \caption{The average sum rate versus the scanning fraction for random selection with pilot power of $4.77$ dBm and BS transmit power of $20$ dBm, with 16 antenna BS serving 10 users, with error bars representing $\pm$ SE.}
    \label{mean_scanning}
\end{figure*}

\subsection{Effect of Partial Scanning}\label{sec:partial_scanning}

Fig.~\ref{mean_scanning} shows the average sum rate versus the scanning fraction with GA-based and LS-based channel acquisition. The fixed-fraction policies $\pi_{\mathrm{unif}}$ and $\pi_{\mathrm{rand}}$ are curves in $\rho_{\mathrm{scan}}$. The adaptive policy (Algorithm~\ref{alg:adaptive_scan}) $\pi_{\mathrm{adapt}}$ ($\eta=0.01$) selects its own scanning fraction, so it is shown as the average rate along its scan order, with a star at its average stopping point. The sparse and compact UPAs do not depend on $\rho_{\mathrm{scan}}$ and appear as horizontal references, evaluated with the same acquisition method as the MA in each panel.

\textbf{Fixed fractions.} At full-grid scanning, the optimized MA with GA estimation gives $\approx 27\%$ higher sum rate than that of LS. Both methods exceed the sparse UPA by wide margins, i.e., by factors of $4.7\times$ and $5.4\times$. The scanning strategies differ in how much scanning they need to approach the full-grid rate. Under GA, $\pi_{\mathrm{unif}}$ saturates at $\rho_{\mathrm{scan}}\approx 40\%$, whereas $\pi_{\mathrm{rand}}$ is still $\approx 9\%$ below the full-grid rate at $\rho_{\mathrm{scan}}=80\%$ and reaches it only at $\rho_{\mathrm{scan}}=1$. In contrast, under LS estimation, the two policies are nearly indistinguishable beyond $\rho_{\mathrm{scan}}\approx 40\%$. Hence, under GA, the random policy is far more sensitive to which locations are measured, whereas under LS, the choice of locations matters mainly at small $\rho_{\mathrm{scan}}$. Compared with the sparse UPA, uniform scanning at the smallest fraction ($\rho_{\mathrm{scan}}=0.0166$) only matches it under both GA and LS, since the few scanned locations are too coarse to be exploited. Random scanning already exceeds it by $3.2\times$ (GA) and $4.0\times$ (LS) at the same fraction.

\textbf{Adaptive scanning.} With $\eta=0.01$, the adaptive policy stops after scanning on average $\bar\rho_{\rm eff}=11.1\%$ of the candidate locations ($\approx107$ of $961$) in $7$ batches, compared with $61$ for a full scan. Under GA, it reaches $99.5\%$ of the full-grid rate, which is $4.6\times$ the sparse UPA. At $\rho_{\mathrm{scan}}\approx0.1$, the adaptive policy is $60\%$ and $25\%$ higher than uniform and random scanning, respectively. It also remains above both fixed policies at the smallest fraction. Under LS, the adaptive policy reaches $93.5\%$ of the full-grid rate, $5.0\times$ the sparse UPA, and $58\%$ and $6\%$ above uniform and random scanning at $\rho_{\mathrm{scan}}\approx0.1$. The difference between the two methods comes from the shape of the trajectory: under GA the rate is nearly flat beyond $\rho_{\mathrm{scan}}\approx7$--$10\%$, so stopping early costs little, whereas under LS it keeps rising slowly up to $\rho_{\mathrm{scan}}\approx40\%$. In summary, the benefit of movable antennas under partial scanning depends on the acquisition method and the scanning budget. GA attains the highest rate, and the adaptive policy retains $99.5\%$ of the full-grid GA rate with about $11\%$ of the scanning locations, far above the sparse UPA.

\begin{figure}[t!]
        \centering         {\includegraphics[width=0.8\textwidth]{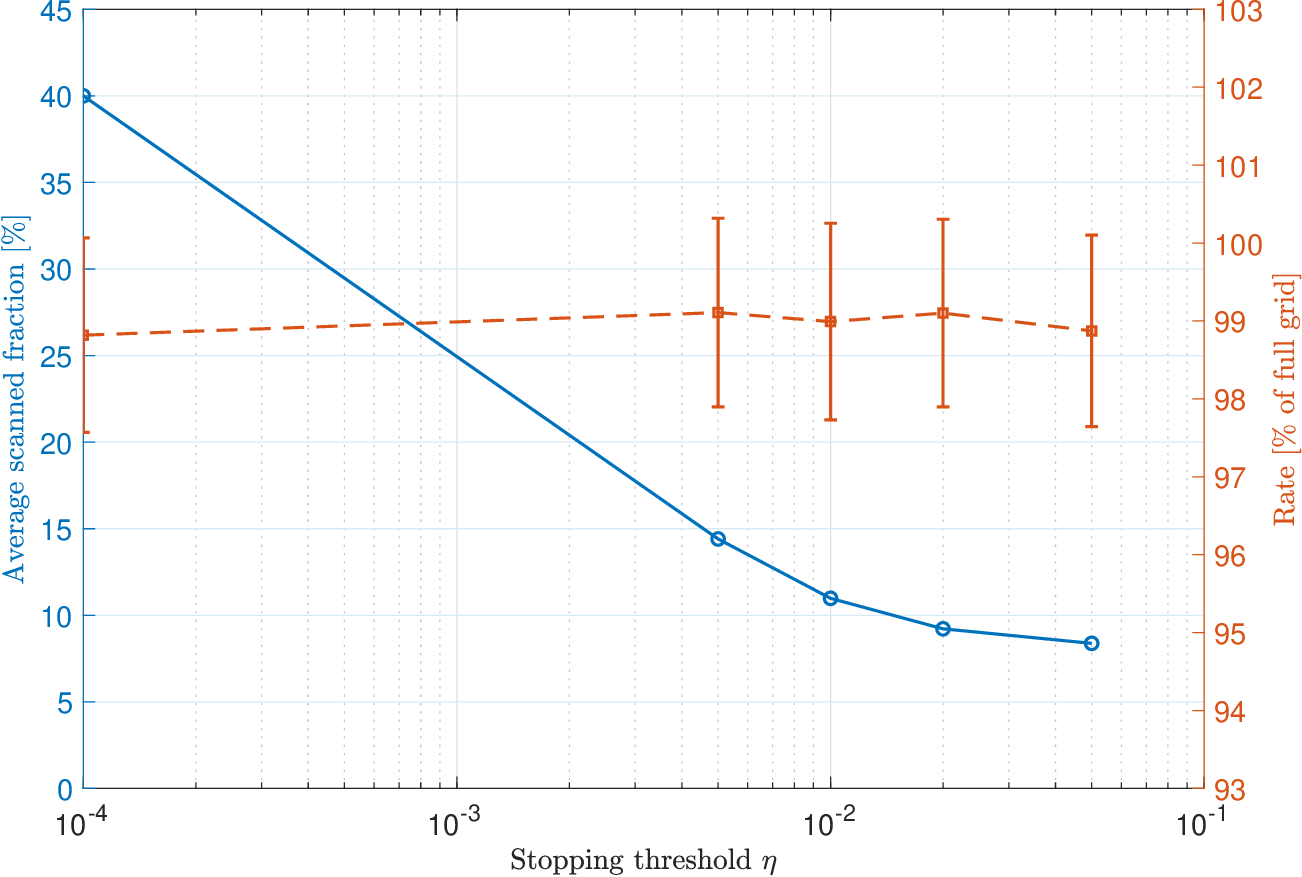}}
        \caption{Stopping behavior and batch-size sensitivity, where error bars represent $\pm$ SE.}
\label{fig:adaptive_efficiency}  \vspace{-2mm}
\end{figure} 

Fig.~\ref{fig:adaptive_efficiency} shows how the
stopping threshold $\eta$ controls the scanning overhead of $\pi_{\mathrm{adapt}}$ ($M_{\mathrm{RF}}=16$, $n_{\min}=2$, patience $4$) in GA estimation. Since the average sum rate saturates quickly under GA, increasing $\eta$ from $10^{-4}$ to $0.05$ reduces the average scanned fraction from $40.0\%$ to $8.4\%$, while the sum rate stays between $98.8\%$ and $99.1\%$ of the full-grid rate. We therefore use $\eta=0.01$, which utilizes $11\%$ of the locations to reach $99.5\%$ of the full-grid rate. 

\subsection{Effect of User Mobility}\label{sec:mobility}

The previous subsection considered a static channel, where $\pi_{\mathrm{adapt}}$ starts from scratch and needs only a fraction of the candidate locations to reach nearly the full-grid rate. When users move, previously acquired CSI becomes outdated and the previously optimized antenna geometry may become suboptimal. Yet, the large-scale propagation geometry varies gradually. We therefore investigate how much scanning is needed to recover the rate after a displacement, and whether re-scanning from a fixed low-cost array geometry is sufficient compared with a full scan or with keeping the previously optimized antennas.

Each $i$th Monte Carlo drop fixes an initial user configuration $\mathbf{B}_{\mathrm{orig}}^{(i)}$. For every $k$th user, a direction $\psi_k\sim\mathcal{U}[-\pi/2,\pi/2)$ is drawn once per drop, and the displaced positions are $\mathbf{B}_{\mathrm{moved}}(d)=\mathbf{B}_{\mathrm{orig}}+d\,[\cos\psi_k;\,\sin\psi_k]$, clipped so that $|\phi_k|\le\pi/2-0.02$, which keeps users in the forward half-space of the array. All displacement values $d$ in a drop therefore lie on the same ray, so the effect of $d$ is not confounded with a change of direction. Averaging over $100$ drops covers many user positions and displacement directions.

\begin{figure*}[t!]
    \centering
    \subfloat[User geometry before and after displacement ($K=10$, one drop).]{ \includegraphics[width=0.3\textwidth]{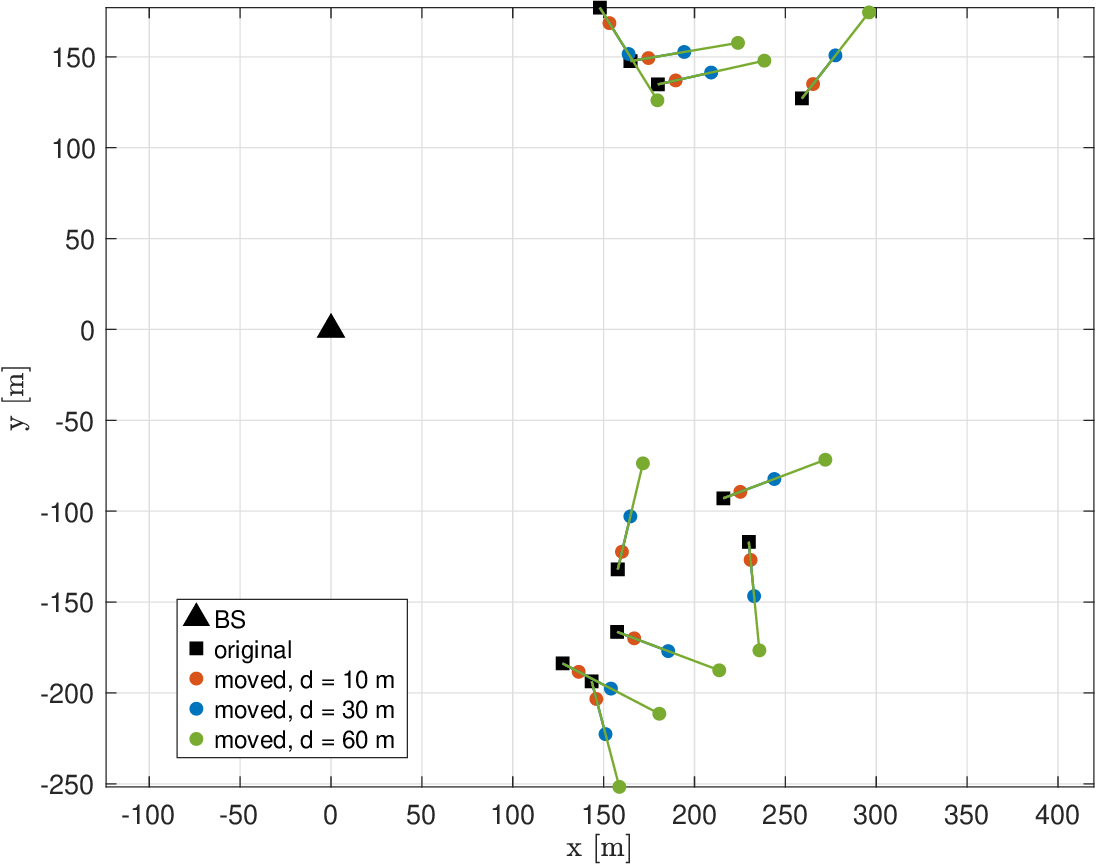}
        \label{user_mob}} \hfill
    \subfloat[Average sum rate versus displacement $d$.]{ \includegraphics[width=0.325\textwidth]{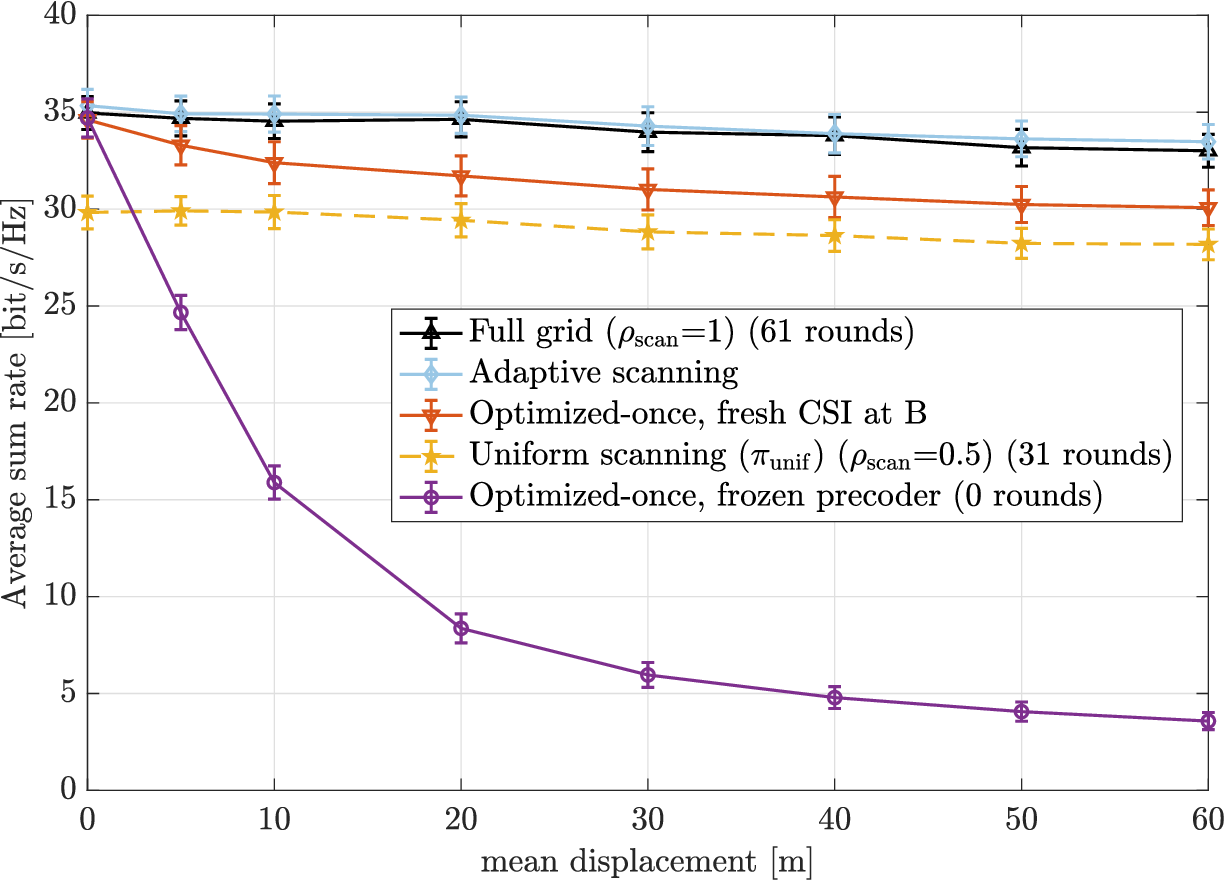}
        \label{rate_usermob} }
    \hfill
    \subfloat[Antenna-scan budget per strategy at the displaced geometry.]{ \includegraphics[width=0.325\textwidth]{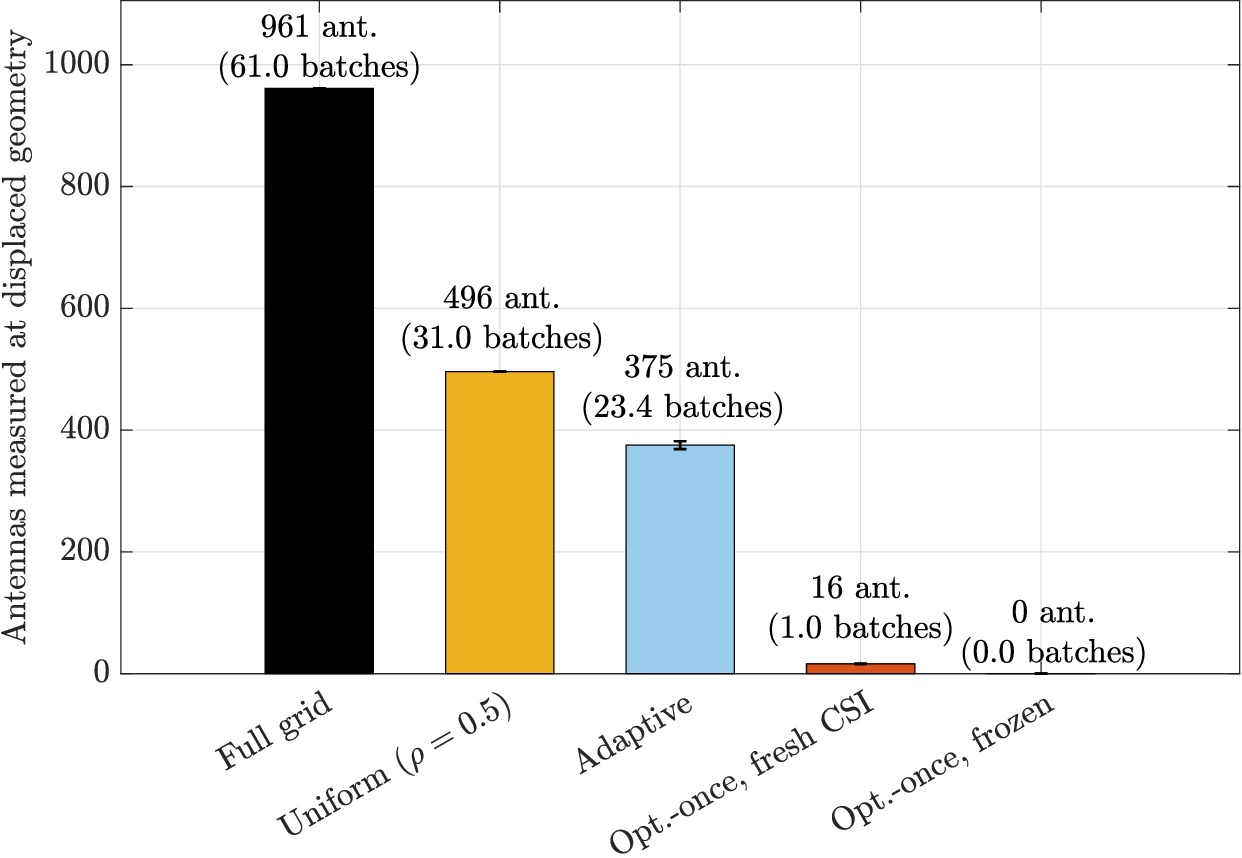}
        \label{fig:mob_budget}}    
        \caption{Effect of user mobility: (a)~displaced geometry for one Monte Carlo drop; (b)~achievable sum rate versus displacement; (c)~antenna-scan budget at the displaced geometry. Frozen CSI reuses the reference precoder with no new pilots; fresh CSI re-estimates the channel at the fixed array; scanning strategies re-select the antenna subset after the move.}
    \label{fig:mobility_panel}
\end{figure*}

For each drop and each $d$, the true channel $\mathbf{H}(\mathbf{B}_{\mathrm{moved}}(d))$ is held fixed. Each strategy averages its rate over $5$ independent pilot realizations. The sum rate is always evaluated on the \emph{true} channel at the displaced positions, while each strategy builds its precoder from its own channel estimate. All strategies therefore face the same physical channel, and the curves differ only in (i) which antennas are active, (ii) whether CSI is refreshed after the move, and (iii) how many pilot rounds are spent. Fig.~\ref{fig:mobility_panel} shows the displaced user geometry for one drop (a), the average sum rate versus displacement (b), and the antenna-scan budget of each strategy (c). We compare the following strategies:
\begin{itemize}
  \item \textbf{Adaptive scanning:} $\pi_{\mathrm{adapt}}$ starts from staggered UPA baseline with $M_{\mathrm{RF}}=16$ antennas and no information about the old geometry. Measurements guide the search, and scanning stops once the patience criterion is met, subject to minimum and maximum scanning budgets.
  \item \textbf{Full-grid scanning, $\rho_{\mathrm{scan}}=1$ (61 rounds):} The whole candidate grid is scanned after the move before antenna selection and precoding. This is the full-scanning benchmark.
  \item \textbf{Uniform scanning, $\rho_{\mathrm{scan}}=0.5$ (31 rounds):} Half of the candidate grid is scanned before selecting the antennas. This is the fixed-budget reference for partial spatial sampling.
  \item \textbf{Optimized-once:} A full-grid scan is run \emph{once} at $\mathbf{B}_{\mathrm{orig}}$ (distance from which the user starts to move), and the selected $M_{\mathrm{RF}}$ antennas are kept after the move. We report two variants: the \emph{frozen precoder} computed at $\mathbf{B}_{\mathrm{orig}}$ (0 rounds) and \emph{fresh CSI} on the same subset (1 round). Comparing the two separates the loss from stale CSI from the loss from an outdated antenna subset.
\end{itemize}

Fig.~\ref{rate_usermob} shows the average sum rate versus $d$. At $d=0$ the frozen and fresh-CSI variants of the optimized-once antenna geometry coincide, as expected, since the channel is identical to the one used for optimization. The curves separate as soon as the users move and fall into two groups.

\emph{Strategies that refresh CSI} degrade slowly. Over the sweep from $d=0$ to $60$~m, full-grid scanning decreases approximately $5.7\%$ and adaptive scanning also shows similar performance as full scan with a loss of only $\approx 5.6\%$ as the user moves. The adaptive curve stays within the confidence intervals of the full-grid curve, so an adaptive scan recovers the full-scan rate without any knowledge of the previous geometry. The optimized-once antenna array with fresh CSI matches full-grid scanning at $d=0$ but shows a decrease of $13\%$ at $60$~m, and the gap widens with $d$. This shows that an antenna subset optimized for the old user positions becomes progressively less suitable as users move. Uniform scanning with $\rho_{\mathrm{scan}}=0.5$ is the lowest of this group, about $5$~bit/s/Hz below the full grid, even though it scans half of the candidate locations. Partial sampling without adaptivity therefore does not reliably identify a good subset.

\emph{The optimized-once frozen precoder} collapses quickly. Its rate drops by $\approx 29\%$ at $d=5$~m, $55\%$ at $10$~m, and almost $90\%$ at $60$~m below the $d=0$ value, so most of the loss occurs within the first few meters. The antenna geometry is identical to that of the fresh-CSI variant, so the gap between the two ($\approx 9$~bit/s/Hz at $5$~m and $\approx 26$~bit/s/Hz at $60$~m) is caused entirely by outdated CSI: the precoder's beams and nulls are misaligned with the new channel. Refreshing the CSI therefore restores most of the lost rate, while re-selecting the antennas adds a smaller further gain.

Fig.~\ref{fig:mob_budget} compares the scanning effort at the displaced geometry. Full-grid scanning measures all $961$ candidate antennas ($61$ batches), and uniform scanning measures about half ($31$ batches). The adaptive policy ($\eta=0.01$) stops after $23.4$ batches on average, i.e., $375$ antennas ($39\%$ of the grid), while achieving the same rate as the full grid. The optimized-once strategies require no scan of the displaced geometry, since their subset was selected once at the old positions: the frozen variant uses $0$ pilot rounds, and the fresh-CSI variant uses a single round on its $16$ antennas. Hence, a single pilot round after the move recovers most of the lost rate, and re-selecting the antennas adds a further $0.7$-$3.4$~bit/s/Hz over the fresh-CSI subset (which grows with $d$). Adaptive scanning provides this gain at a small fraction of the full-scan cost. Together, the results quantify the trade-off between CSI freshness, antenna-position adaptation, and pilot overhead: refreshing CSI is essential, while re-adapting the antenna positions further improves the rate at a moderate scanning cost.

\section{Conclusion}\label{Sect_conclusion}

This paper developed a framework for discrete movable-antenna (MA) selection under realistic channel estimation. It combines a geometry-aware (GA) codebook for channel estimation, an energy-based adaptive policy that decides how many candidate positions to probe, and an evaluation of the resulting scanning cost for static and mobile users. The central finding is that the benefit of MA selection depends as much on the method used to acquire CSI as on the antenna positions themselves. Evaluating a candidate position requires probing it, so the scanning cost grows with the size of the candidate region and can erase the benefit of reconfigurability. We therefore studied how much of the region must be scanned, how the scanned positions should be chosen, and how the answer changes when users move. The main takeaways are as follows:

\begin{enumerate}
  \item \emph{Movable antennas deliver a considerable gain over fixed arrays, provided the channel estimation is matched to the antenna geometry.} With perfect CSI, the optimized MA achieves about $5.3\times$ the rate of the sparse UPA in the considered simulation setup. With practical estimation, the GA codebook retains $78.5\%$ of this rate and improves on the geometry-agnostic alternatives by $6.5\%$ to $27.0\%$. A well-designed codebook therefore reduces the estimation error and preserves a large part of the MA gain.

  \item \emph{Probing all candidate positions is unnecessary.} The proposed adaptive policy reaches $99.5\%$ of the full-grid GA rate after probing about $11\%$ of the candidate positions, a $9\times$ reduction in probing cost relative to a full scan. The best fixed-fraction policy needs about $40\%$ coverage to reach the same level, i.e., about $3.6\times$ the adaptive cost.

  \item {Under user mobility, refreshing the CSI is more valuable than re-optimizing the antenna positions.} Adaptive scanning stays within about $6\%$ of its static rate when users move up to $60$~m from their original positions, and it matches the full-grid scan. If the antenna positions are optimized only once, at the original user positions, and only the CSI is refreshed afterward, the rate drops by about $13\%$ at $60$~m, yet it remains well above both frozen CSI and uniform half-grid scanning. Re-selecting the antennas adds about $0.7$ to $3.4$~bit/s/Hz at the cost of a new scan. Given this added complexity, the freshness of the CSI matters more than continuous antenna repositioning.

\end{enumerate}
Many existing MA studies assume that the channel is known at every candidate position. Our results show that this assumption is neither free nor necessary: the practical value of a movable-antenna system is determined by how efficiently the channel at the candidate positions can be learned, and a geometry-aware codebook combined with adaptive scanning makes this affordable. The proposed acquisition model, scanning policies, and mobility evaluation provide a good starting point for comparing new codebooks and scanning rules, and for extending the analysis beyond free-space LoS to multipath channels and to the joint design of codebook resolution and scanning policy.

\balance
\bibliographystyle{IEEEtran}
\bibliography{IEEEabrv,mybib}

\end{document}